\documentclass[reprint,showpacs,12pt,a4paper,final,amsmath,amssymb]{iopart}
\usepackage{iopams}  
\usepackage[breaklinks=true,colorlinks=true,linkcolor=blue,urlcolor=blue,citecolor=blue]{hyperref}
\usepackage{graphicx}% Include figure files
\usepackage{dcolumn}% Align table columns on decimal point
\usepackage{bm}% bold math

\def\be{\begin{equation}}
\def\ee{\end{equation}}
\def\bea{\begin{eqnarray}}
\def\eea{\end{eqnarray}}

\def\no{\nonumber}
\def\ra{\rangle}
\def\la{\langle}
\def\s{\sigma}
\def\t{\tau}

\usepackage{color}

\begin{document}

\title[Mixed-spin system with supersolid phases]{Mixed-spin system with supersolid phases:\\ Magnetocaloric effect and thermal properties}

\author{F. Heydarinasab$^{1,2}$ and J. Abouie$^1$}
\address{$^1$Department of Physics, Institute for Advanced
Studies in Basic Sciences (IASBS), Zanjan 45137-66731, Iran}
\address{$^2$Department of Physics, Faculty of Science,
University of Sistan and Baluchestan, Zahedan, Iran}
%\email[]{fheydari@iasbs.ac.ir} \email[]{jahan@iasbs.ac.ir}

\begin{abstract}
Recently, it has been shown that two dimensional frustrated mixed-spin systems with anisotropic exchange interactions display supersolid phases in their ground state phase diagrams even in the absence of long-range interactions. 
In this paper, using cluster mean field theory, we investigate the effects of thermal fluctuations on the ground state phases of this kind of systems and show that various thermal solids and thermal insulators emerge around the ground state solid and Mott insulating phases. 
We also study the thermodynamic properties and magnetocaloric effect of these systems and demonstrate that at low temperatures, a large cooling rate is seen in the vicinity of the solid-supersolid, solid-superfluid and Mott insulator-superfluid critical points, with the large accumulation of the entropy and the minimums of the isentropes. Our results show the sign change of the magnetocaloric parameter inside the solids and the Mott insulator, which is a characteristic of ordered phases.
\end{abstract}

\date{\today}

%\maketitle

%%%%%%%%%%%%%%%%%%%%%%%%%%%%%%%%%%%%%%%%%%%%%%%%%%Introduction %%%%%%%%%%%%%%%%%%%%%%%%%%%%%%%%%%%%%%%%%%%%%%%%%%

\submitto{\JPCM}
%\maketitle

%%%%%%%%%%%%%%%%%%%%%%%%%%%%%%%%%%%%%%%%%%%%%%%%%%Introduction %%%%%%%%%%%%%%%%%%%%%%%%%%%%%%%%%%%%%%%%%%%%%%%%%%
\section{Introduction}\label{sec:intro}

Supersolid is an exotic state of matter, characterizing by the coexistence of solid and superfluid long-range orders \cite{thouless1969flow, andreev1971quantum,Matsuda701, Liu1973}. Combination of these two apparently antithetical properties has attracted the attentions of both experimentalists and theorists, and searching for this phase has become one of the main subjects of condensed matter and cold atoms physics \cite{kim2004probable, kim2004observation, li2017stripe, leonard2017supersolid, tanzi2019observation, bottcher2019transient, chomaz2019long, natale2019excitation}. 
An appropriate ground for searching various supersolid phases are quantum spin systems \cite{sengupta2007spin, peters2009spin,rossini2011spin,PhysRevLett.97.127204, chen2010field, laflorencie2007quantum, PhysRevLett.100.090401,guo2014stripe, thomson2015bose, murakami2013supersolid, albuquerque2011phase, wierschem2013columnar, picon2008mechanisms, ng2017field, momoi2000magnetization,sengupta2008ground, toth2012competition, su2014magnetic, sengupta2007field,ueda2013nematic, selke2013multicritical, morita2019magnetization}. 
Recently it has been shown experimentally that mixed-spin systems, composed of two kinds of spin, display various supersolid phases in their ground state phase diagrams \cite{tsurkan2017ultra, ruff2019multiferroic}. 
Mixed-spin systems are a special class of spin models, where their universality class is completely different from uniform spin models \cite{trumper2001antiferromagnetically, PhysRevB.70.184416, PhysRevB.73.014411, 1742-5468-2011-08-P08001, fukushima2004thermodynamic, solano2010magnetic, solano2011entanglement, karchev2008towards}. 
We recently have obtained the ground state phase diagrams of two mixed-spin systems on the square lattice with two different arrangements and demonstrated theoretically that aside from solid, superfluid and Mott insulating phases, they possess various supersolid phases in their ground state phase diagrams, even in the absence of long-range interactions \cite{heydarinasab2017inhomogeneous, heydarinasab2018spin}.

%Recent progress provided the ability of manufacturing these kind of models on the optical lattices \cite{safaei2013raman, saugmann2019magnetic}. 

In this paper, in the first part we investigate the effects of thermal fluctuations on the stability of the ground state phases of the mixed-spin systems. Using cluster mean field theory (CMFT), we show that, in comparison with the off-diagonal superfluid order, the diagonal solid orders are more stable against thermal fluctuations.
We demonstrate that various thermal solids and thermal insulators also emerge around the ground state solid and Mott insulating
phases. We also show that the solid-solid and supersolid-Mott
insulator phase transitions in these systems maintain first order
even at high temperatures where the ground state phases around these
transition points are washed out completely.

In the second part of this paper, we study thermodynamic properties
and magnetocaloric effect (MCE) of the models. MCE, introduced by
Warburg \cite{warburg1881magnetische}, is the temperature variations
of the magnetic systems in response to the adiabatic changes of
magnetic field. In general, due to the accumulation of entropy in the
vicinity of the transitions \cite{zhitomirsky2003enhanced, garst2005sign, boyarchenkov2007quantum}, MCE highly enhances near
the quantum phase transitions \cite{garst2005sign,  boyarchenkov2007quantum, zhu2003universally, zhitomirsky2004magnetocaloric, sosin2005magnetocaloric, lang2010large, wolf2011magnetocaloric, schmidt2007magnetocaloric, strassel2015magnetocaloric,  ohanyan2012magnetothermal, galisova2016reentrant, andreenko1989magnetocaloric, seabra2009two, wolf2014cooling}, so it would be an empirical quantity for measuring experimental phase diagram of different systems \cite{bianchi2002first, jaime2004magnetic, silhanek2006irreversible, samulon2009asymmetric, kohama2012anisotropic, kohama2014entropy, sun2018low, bocarsly2018magnetoentropic, wang2018quantum}. 
Aside form the
fundamental interests, the magnetocaloric effect has great importance for magnetic cooling techniques.
% which has been employed over the years to reach temperatures in the sub-kelvin range\cite{lounasmaa1974experimental}.
Certain progress has also been achieved to utilize this technique for room temperature refrigeration \cite{pecharsky1997giant, tegus2002transition, pecharsky1999magnetocaloric, tishin1999magnetocaloric, pereira2009magnetocaloric, samanta2012giant, choudhury2014tuning, zhang2018enhanced, huang2014giant}. 
Different parameters affect the cooling rate. For example it has been shown that the higher the density of the magnetic moments and their spin number, the greater the cooling power of a refrigerant is \cite{zhitomirsky2003enhanced}. Also, residual entropy in the frustrated spin systems results the larger cooling rate \cite{zhitomirsky2003enhanced, zhitomirsky2004magnetocaloric, sosin2005magnetocaloric,  pereira2009magnetocaloric, schnack2007enhanced, honecker2006magnetocaloric, honecker2009magneto, derzhko2006universal, richter2009heisenberg}. Moreover it is known that the magnetocaloric effect is quite large in ferrimagnetic materials \cite{boyarchenkov2007quantum, clark1969cooling, luo2009observation, kumar2018magnetocaloric}. 

In this paper we study the MCE in the two different frustrated mixed-spin systems on the square lattice. We demonstrate that at low temperatures, a large cooling rate is seen in the vicinity of the solid-supersolid, solid-superfluid and Mott insulator-superfluid quantum critical points, with the large accumulation of the entropy and the minimums of the isentropes. Up to our knowledge, this is the first study about the MCE in the supersolid phases, which a large cooling rate around this phase in addition to the multi-peak structure of the specific heat could be a signature of this phase. 

This paper is organized as follows. In Sec. \ref{sec:model}, we
introduce our frustrated mixed-spin models on the square lattice
with two different arrangements. In Sec. \ref{sec:phase-diagrams}, we
briefly review the CMFT ground state phase diagrams of the introduced models,
and investigate the effects of thermal fluctuations on the stability
of the ground state phases. In this section we also present the
temperature phase diagram of the systems.  In Sec.
\ref{Thermodynamic-functions}, we study the isothermal and also temperature variations
of different thermodynamic functions, such as magnetization,
magnetic susceptibility, specific heat and entropy. The
magnetocaloric effects in different phases are also investigated in
this section. Finally, we will summarize our results and give the
concluding remarks in Sec. \ref{sec:summary}.

%%%%%%%%%%%%%%%%%%%%%%%%%%%%%%%%%%%%%%%%%%%%%%% %%%%%%%%%%%%%%%%%%%%%%%%%%%%%%%%%%%%%%%%%%%%%%%%%%%
\section{Mixed spin-(1, 1/2) system with different arrangements}\label{sec:model}
%%%%%%%%%%%%%%%%%%%%%%%%%% FIG 0 %%%%%%%%%%%%%%%%%%
\begin{figure}[ht!]
\centering
\includegraphics[width=80mm]{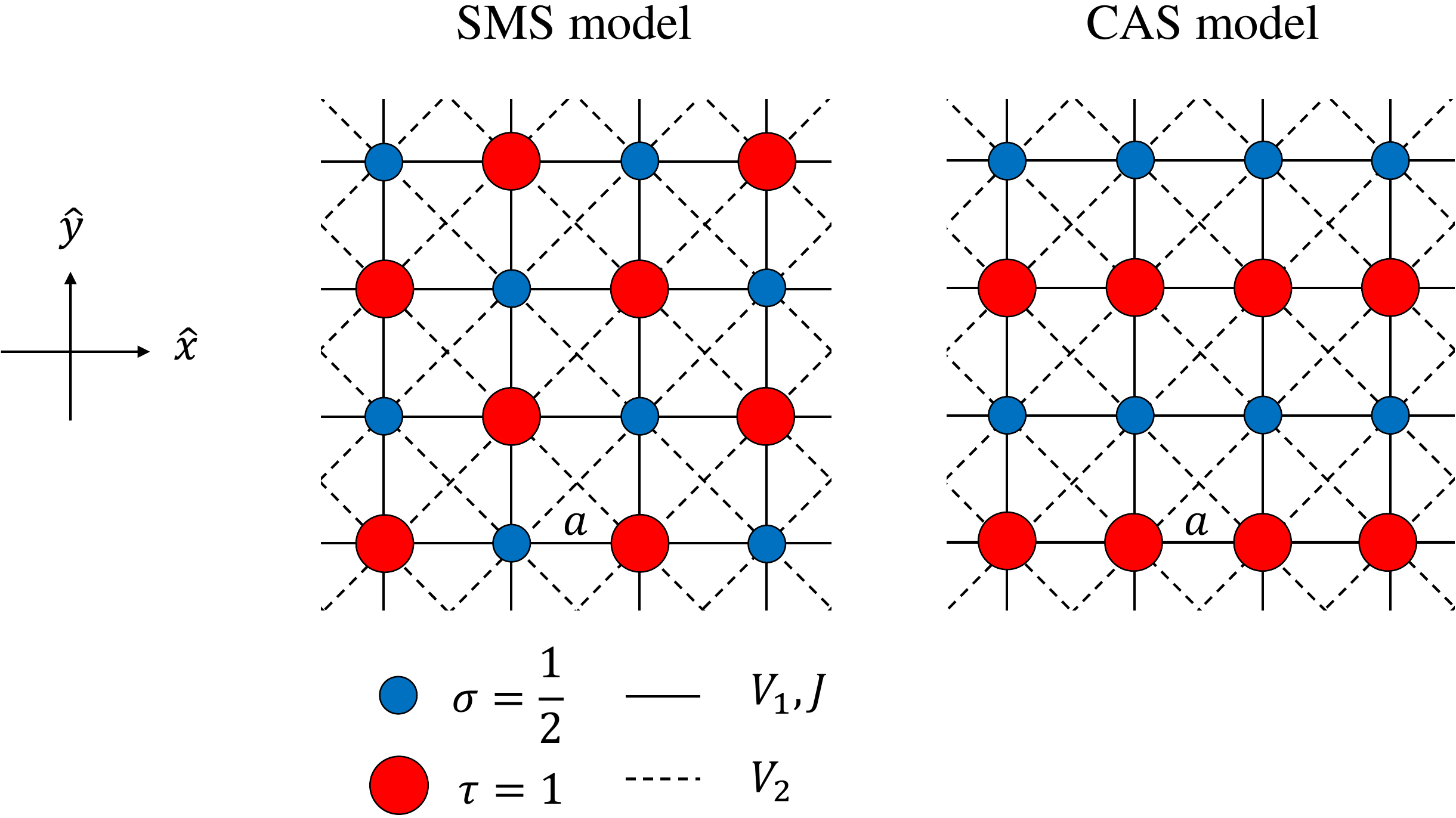}
\caption{(Color online) Schematic illustration of the mixed spin
system with different arrangements. Left: mixed spin system with a
staggered sequences (SMS model), and right: alternatively coupled
spin chains (CAS model). The small and large circles denote the spin
1/2 and 1, respectively. The solid and dotted lines are nearest
neighbor and next nearest neighbor interactions, respectively. $a$
is the lattice constant.}\label{fig:model}
\end{figure}
%%%%%%%%%%%%%%%%%%%%%%%%%%%%%%%%%%%%%%%%%%%%%%%%
We consider a two-dimensional mixed spin system, composed of two
different spins: $\t=1$ and $\s=1/2$, interacting via the following
Hamiltonian:
\begin{equation}
H=H_{\s}+H_{\t}+H_{\s\t}, \label{Hamiltonian}
\end{equation}
where $H_\s$, $H_\t$, and $H_{\s\t}$ include the interactions
between spins $\s$, spins $\t$, and spins $\s$ and $\t$,
respectively. Depending on the sequences of the spins $\s$ and $\t$
on the square lattice, two different arrangements are considered in
this paper. The first is an staggered mixed spin system where each
spin-1/2 is surrounded by four spins $\t$ (SMS model) and the second
is an stripe mixed spin system where spin-$\s$ chains are
alternatively coupled to spin-$\t$ chains (coupled alternating spin
chains (CAS) model). These two systems are schematically illustrated
in Fig. \ref{fig:model}.

The Hamiltonians $H_\s$, $H_\t$, and $H_{\s\t}$ for SMS and CAS
models are respectively given by
\begin{eqnarray}
\nonumber H_\s&=& V_2 \sum_{nnn}\s_i^z \s_j^z-h\sum_i \s_i^z,\\
\nonumber H_\t&=& V_2 \sum_{nnn}\t_i^z \t_j^z-h\sum_i \t_i^z,\\
H_{\s\t}&=&\sum_{nn}[-2J(\s_i^x \t_j^x+\s_i^y \t_j^y)+V_1\s_i^z
\t_j^z], \label{eq:SMS-Hamiltonian}
\end{eqnarray}
and
\begin{eqnarray}
\nonumber H_\s&=&\sum_{nn}[-2J(\s_i^x \s_j^x+\s_i^y \s_j^y)+V_1\s_i^z \s_j^z]-h\sum_i \s_i^z,\\
\nonumber H_\t&=&\sum_{nn}[-2J(\t_i^x \t_j^x+\t_i^y \t_j^y)+V_1\t_i^z \t_j^z]-h\sum_i \t_i^z,\\
\nonumber H_{\s\t}&=&\sum_{nn}[-2J(\s_i^x \t_j^x+\s_i^y
\t_j^y)+V_1\s_i^z \t_j^z]+V_2\sum_{nnn}\s_i^z\t_j^z,\\
\label{eq:CAS-Hamiltonian}
\end{eqnarray}
where the summations $\sum_{nn}$ and $\sum_{nnn}$ run over nearest
neighbor (NN) and next nearest neighbor (NNN) sites, $V_1$ and $V_2$
are respectively the NN and NNN interactions, and $h$ is a magnetic
field along $z$ direction. These mixed (1/2,1) spins models correspond to modified Bose-Hubbard models with respectively hard-core and semi hard-core bosons with occupancy up to one and two particles per lattice sites \cite{heydarinasab2017inhomogeneous,heydarinasab2018spin}.

Both spin models possess the rotational U(1) symmetry as well as the
discrete translational symmetry of the square lattice. The
translational vectors in the SMS and CAS lattices are respectively
$2a\hat{x}+2a\hat{y}$ and $a\hat{x}+2a\hat{y}$, where $a$ is the
lattice constant, as shown in Fig. \ref{fig:model}. According to the
spontaneously breaking of one or both of these symmetries various
first- and second-order phase transitions occur and different
diagonal and off-diagonal long-range orders appear in these systems.
In the following section we will briefly review the ground state
phases of the models (\ref{eq:SMS-Hamiltonian}) and
(\ref{eq:CAS-Hamiltonian}), which are presented in Refs.
\cite{heydarinasab2017inhomogeneous,heydarinasab2018spin}, and then
study the effects of thermal fluctuations on the stability of the
ground state phases and obtain the temperature phase diagrams of the
SMS and CAS models.

%%%%%%%%%%%%%%%%%%%%%%%%%%%%%%%%%%%%%%%%%%%%
\section{Phase diagrams}\label{sec:phase-diagrams}
%%%%%%%%%%%%%%%%%%%%%%%%%%%%%%%%%%%%%%%%%%%%

Recently, we have studied the ground state properties of the SMS and
the CAS models, using different methods like mean field approximation,
cluster mean field theory and linear spin wave approach, and shown
that various solids, supersolids, and Mott insulator emerge in their
ground state phase
diagrams \cite{heydarinasab2017inhomogeneous,heydarinasab2018spin}. Below, first we will briefly review the zero temperature properties
of these phases and then obtain the temperature phase diagrams of
the SMS and the CAS models.

%%%%%%%%%%%%%%%%%%%%%%%%%%%%%%%%%%%%%%%%%%%%
\subsection{SMS model}
%%%%%%%%%%%%%%%%%%%%%%%%%%%%%%%%%%%%%%%%%%%%

%%%%%%%%%%%%%%%%%%%%%%%%%% FIG 0 %%%%%%%%%%%%%%%%%%
\begin{figure}[h]
\centering
\includegraphics[width=78mm]{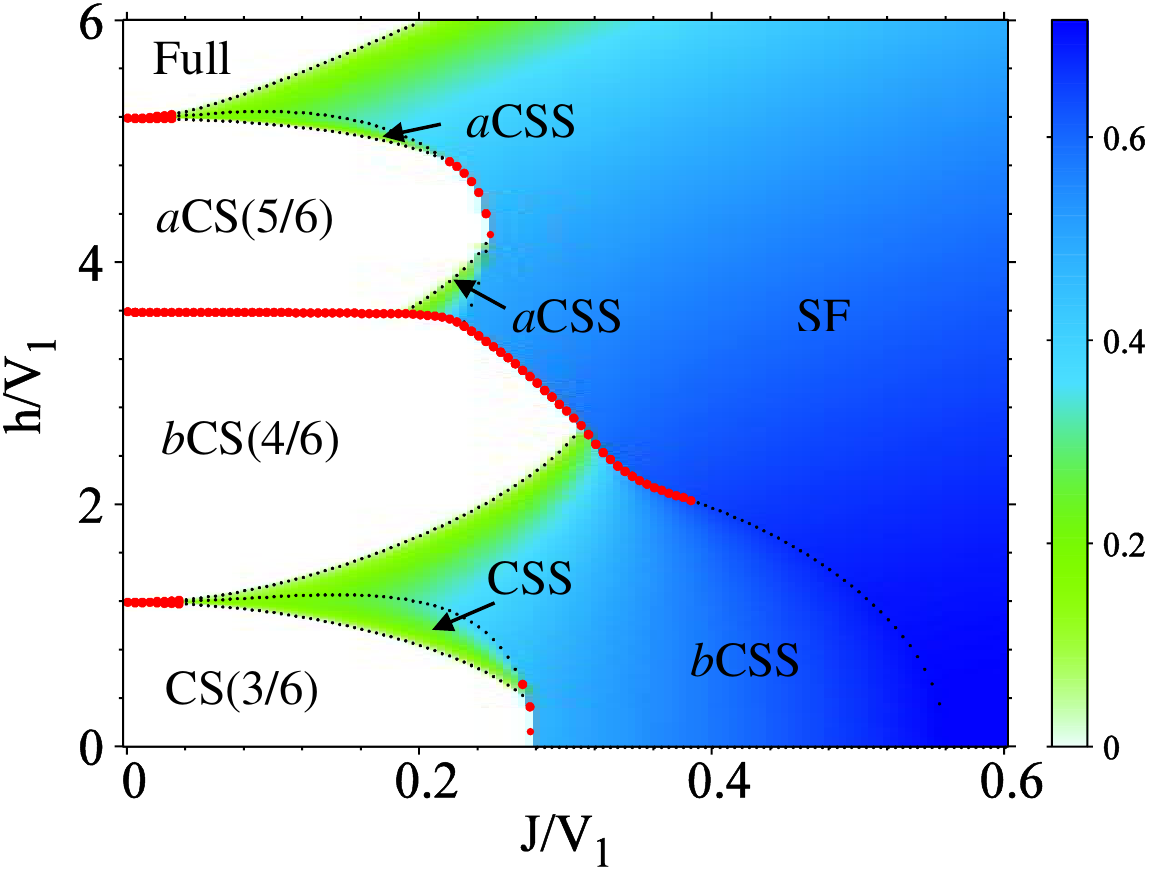}
\includegraphics[width=77mm]{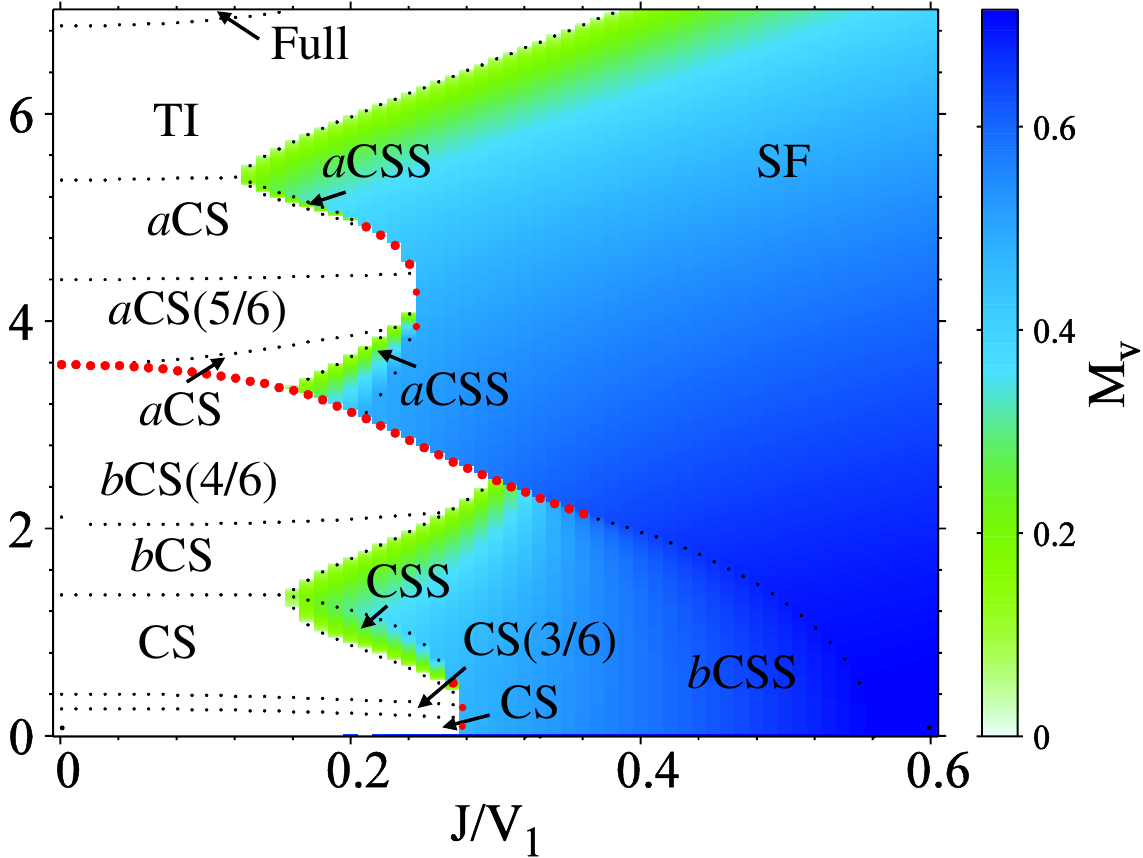}
\caption{(Color online) $J-h$ phase diagrams of the
SMS model for $\frac{V_2}{V_1}=0.6$. Left: at zero temperature, and right: at $T/V_1=0.1$.
The density of colors shows amount of the off-diagonal order parameter:
$M_{v}=((M^x_T)^2 + (M^y_T)^2)^{1/2}$ where $M_T^{x (y)}$ is the
total magnetization in $x (y)$ direction. The red (black) dotted
lines show first-order (second-order) phase transitions.  The
different orders are defined in the table \ref{tab:SMS}.
}\label{fig:SMS-Jh-atT}
\end{figure}
%%%%%%%%%%%%%%%%%%%%%%%%%%%%%%%%%%%%%%%%%%%%%%%%
%%%%%%%%%%%%%%%%%%%%%%%%%%%%%%%%%%%%%%%%%%%%% Definition of the orders  %%%%%%%%%%%%%%%%%%%%%%%%%%%%%%%%%%%%%%%%%%%%%

\begin{table}
\caption{Definitions of various ground state phases of the SMS model. Because of
the NNN interactions we divide the subsystem with spin $\s=1/2$
($\t=1$) into two sublattices A and C (B and
D) \cite{heydarinasab2017inhomogeneous}. According to the relations
between the longitudinal magnetizations of these sublattices,
different phases appear in the ground state and temperature phase
diagrams of the SMS model. The longitudinal magnetizations
$m_{A(C)}^z=\la \s_{A(C)}^z \ra$ and $M_{B(D)}^z=\la \t_{B(D)}^z
\ra$ are obtained using cluster mean field theory.}
\label{tab:SMS}
\begin{center}
%\begin{ruledtabular}
\begin{tabular}{ccc}
\br
Phases & sublattices magnetizations & $M_v$\\
\hline
SF& $m_A^z=m_C^z$, ~$M_B^z=M_D^z$& $\neq 0$\\
Full& $m_A^z=m_C^z=1/2$, ~$M_B^z=M_D^z=1$& 0\\
MI(4/6)& $m_A^z=m_C^z$, ~$M_B^z=M_D^z$& 0\\
CS(3/6)& $m_A^z=-m_C^z$, ~$M_B^z=-M_D^z$& 0\\
$b$CS(4/6)& $m_A^z=m_C^z$, ~$M_B^z=-M_D^z$& 0\\
$a$CS(5/6)& $m_A^z=-m_C^z$, ~$M_B^z=M_D^z$& 0\\
CSS& $m_A^z \neq m_C^z$,~ $M_B^z \neq M_D^z$& $\neq 0$\\
$b$CSS& $m_A^z=m_C^z$,~$M_B^z \neq M_D^z$& $\neq 0$\\
$a$CSS& $m_A^z\neq m_C^z$, ~$M_B^z=M_D^z$& $\neq 0$\\
\br
\end{tabular}
%\end{ruledtabular}
\end{center}
\end{table}

%%%%%%%%%%%%%%%%%%%%%%%%%%%%%%%%%%%%%%%%%%%%%
In the left panel of Fig. \ref{fig:SMS-Jh-atT}, we have plotted the
ground state $J-h$ phase diagram of the SMS model for the
frustration value $V_2/V_1=0.6$ \cite{heydarinasab2017inhomogeneous}. Various ground state phases of this model is defined in the table \ref{tab:SMS}. 
For small values of $J/V_1$ (i.e. very small hopping energies), by
increasing magnetic field $h$ various checkerboard solid orders such as CS(3/6),
$b$CS(4/6) and $a$CS(5/6) with different fillings appear in the
ground state phase diagram. In bosonic language, the
fractional numbers 3/6, 4/6, and 5/6 refer to the average number of
bosons on each unit cell \cite{heydarinasab2017inhomogeneous}. For
small magnetic fields, the translational symmetry of both subsystems
with spin $\t$ and $\s$ break spontaneously, and the CS(3/6) solid
appears in the phase diagram. In this phase the spins $\t$ as well
as the spins $\s$ are antiparallel and the average number of bosons
on each unit cell is $3/6$. By increasing of the magnetic field, the
spins $\s$ align parallel to the magnetic field and the
translational symmetry of the subsystem with spin $\s$ is restored,
where a phase transition to the $b$CS(4/6) phase occurs. By more
increasing of the magnetic field, the $b$CS(4/6) changes to the
$a$CS(5/6) solid. In this phase the spins-$\t$ flip to the magnetic
field direction and the translational symmetry of the subsystem with
spin $\t$ is restored, while the translational symmetry is broken in the subsystem with spin $\s$. Finally, at strong magnetic fields both the
translational and the U(1) symmetries are restored and the system
enters to the Mott insulating Full phase, where all spins align
parallel to the magnetic field. An enhancement of the hopping energy
$J/V_1$, adds a superfluid component to these solid orders, and
different supersolid phases such as CSS, $a$CSS and $b$CSS emerge
around the solid phases, where both the translational and U(1)
symmetries are broken. Further increasing of hopping energy,
restores the translational symmetry and the system enters to the
superfluid phase (SF). In the ground state phase diagram of the SMS model
for $\frac{V_2}{V_1}<0.5$ (not shown), a MI(4/6) Mott insulating
phase takes place rather than CS(3/6) and $b$CS(4/6) solids at small
and moderate magnetic fields. In this phase both the translational
and the U(1) symmetries are preserved, and spins-$\t$ ($\s$) are parallel (anti-parallel) to the magnetic field direction.

\begin{figure}[h]
\centerline{\includegraphics[width=80mm]{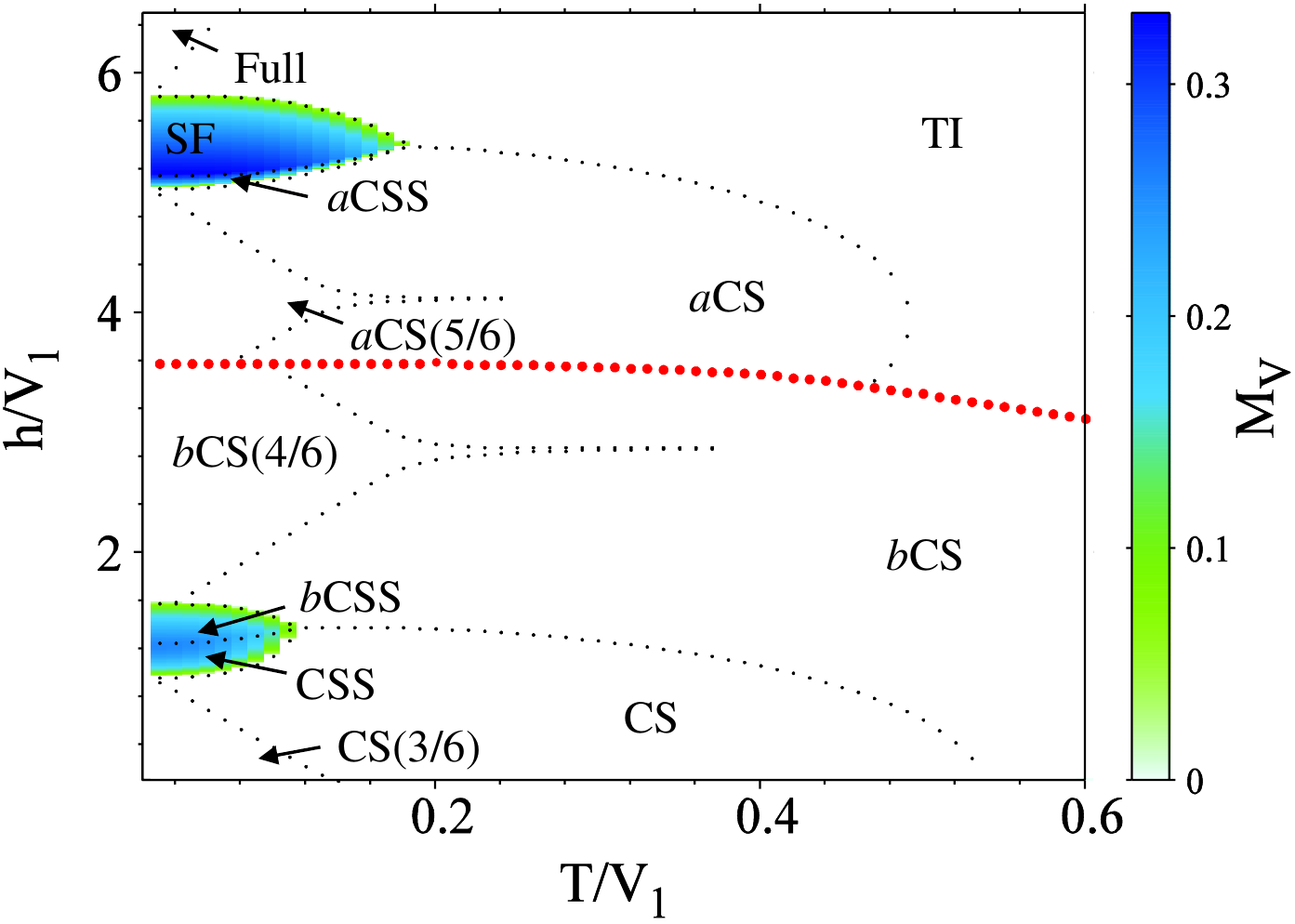}}
\caption{(Color online) $T-h$ phase diagram of the SMS model for
$\frac{V_2}{V_1}=0.6$ at $\frac{J}{V_1}=0.165$ line, where all
phases exist on the ground state phase diagram. The red (black)
dotted lines show first-order (second-order) phase transitions.}
\label{fig:CPD-Th}
\end{figure}

In order to see the effects of thermal fluctuations on the ground
state phases of the SMS model, utilizing cluster mean field theory
we have obtained the $J-h$ phase diagram of the SMS model at
temperature $T/V_1=0.1$. In general, thermal fluctuations melt
plateaus on longitudinal magnetization curve versus magnetic field
and reduce amount of diagonal and off-diagonal order parameters. Therefore, in
the presence of temperature aside from the
ground state phases, several thermal solids like CS, $b$CS and $a$CS
also appear respectively around CS(3/6), $b$CS(4/6) and $a$CS(5/6)
phases, as shown in the right panel of Fig. \ref{fig:SMS-Jh-atT}.
In these thermal solids, in contrast with the ground state solid
phases, longitudinal magnetization varies by magnetic field; whereas the corresponding solid orders persist. In the CS(3/6) the translational symmetry in both subsystems breaks, and the magnetization varies with magnetic field. In the $b$CS ($a$CS) thermal solid phase spins in the subsystem with spin $\t$ ($\s$) are anti-parallel, while the longitudinal magnetization varies with $h$. Actually, since the entropy increases at all the transition points, melting begins from the phase borders and these thermal solids appear
around the ground state solid phases. Plateaus' melting soften the transitions and we expect the transition between solid phases to be mediated by thermal solid orders. However this is not the case for
the $b$CS(4/6)-$a$CS(5/6) transition. This transition remains first
order even though the plateaus melt completely and the $a$CS solid
order is washed out, see Fig. \ref{fig:CPD-Th}. Moreover, in a region below the Full phase, the
SMS model experiences thermal insulator (TI) phase. This phase is a
weak Mott insulator in the sense that it preserves both the
translational and the U(1) symmetries of the original Hamiltonian,
but is different form the ground state Mott insulator since the
magnetization increases gradually with $h$ in the TI phase.
Furthermore, by increasing temperature the regions with the
superfluid and supersolid orders become smaller and finally
disappear at $J/V_1$ around $T/V_1$. For $\frac{V_2}{V_1}<0.5$, the TI phase
also appears around the MI(4/6) phase (not shown).

%%%%%%%%%%%%%%%%%%%%%%%%%% FIG 0 %%%%%%%%%%%%%%%%%%
%%%%%%%%%%%%%%%%%%%%%%%%%%%%%%%%%%%%%%%%%%%%%%%%%

In order to obtain the transition temperatures for different phases, 
we have plotted in Fig. \ref{fig:CPD-Th} the $T-h$ phase diagram of
the SMS model for $\frac{V_2}{V_1}=0.6$, and $\frac{J}{V_1}=0.165$. In the presence of thermal fluctuations in the  solid phases, plateaus'
width on the longitudinal magnetization curve versus magnetic field
decreases gradually and disappears eventually at a transition
temperature which depends on the strength of frustration, the
hopping energy ($J/V_1$) and the magnetic field ($h/V_1$). 

At zero temperature, in the superfluid phase each particle is spread
out over the entire lattice, with long range phase coherence. At
finite temperatures, the superfluid order at the small hopping
energy is suppressed and the system undergoes a transition to the TI
phase where the U(1) symmetry is preserved and the filling factor is
not conserved. The superfluid order at larger hopping energy maintains up to larger temperature. 
For example, when $J/V_1=0.165$, the superfluid phase persists up to $\frac{T}{V_1}=0.18$, while for $J/V_1=0.22$ the transition temperature is $\frac{T}{V_1}=0.33$. 
There are some narrow regions at the lower border of
the SF phase which transform to the $a$CSS supersolid phase. In this phase thermal
fluctuations try to destroy both the diagonal and the off-diagonal long range orders. 
Competing solid and superfluid orders affects transition temperature for the supersolid order. Generally, in comparison with the solid orders the superfluidity order is more
fragile, and by increasing temperature the superfluidity order
destroys at a critical temperature where the supersolid transforms
to a thermal solid. Therefore, CSS and $b$CSS phase phases persist up to  $\frac{T}{V_1}=0.12$ when  $\frac{J}{V_1}=0.165$, while for $\frac{J}{V_1}=0.22$ the transition temperature is $\frac{T}{V_1}=0.25$. Also $a$CSS phases is present up to temperature $\frac{T}{V_1}=0.18$ for  $\frac{J}{V_1}=0.165$, while this phases transforms to $a$CS phase at larger temperature $\frac{T}{V_1}=0.44$ for the $\frac{J}{V_1}=0.22$. Stability of the supersolid phase up to a
temperature comparable with the interaction energy $V_1$, makes the
SMS system a playground for experimental realization of the
different supersolid phases in mixed spin systems.

%%%%%%%%%%%%%%%%%%%%%%%%%%%%%%%%%%%%%%%%%%%%%%%%%%%%%%
\subsection{CAS model}
%%%%%%%%%%%%%%%%%%%%%%%%%%%%%%%%%%%%%%%%%%%%%%%%%%%%%%
%%%%%%%%%%%%%%%%%%%%%%%%%% FIG 0 %%%%%%%%%%%%%%%%%%
\begin{figure}[h]
\centering
\includegraphics[width=78mm]{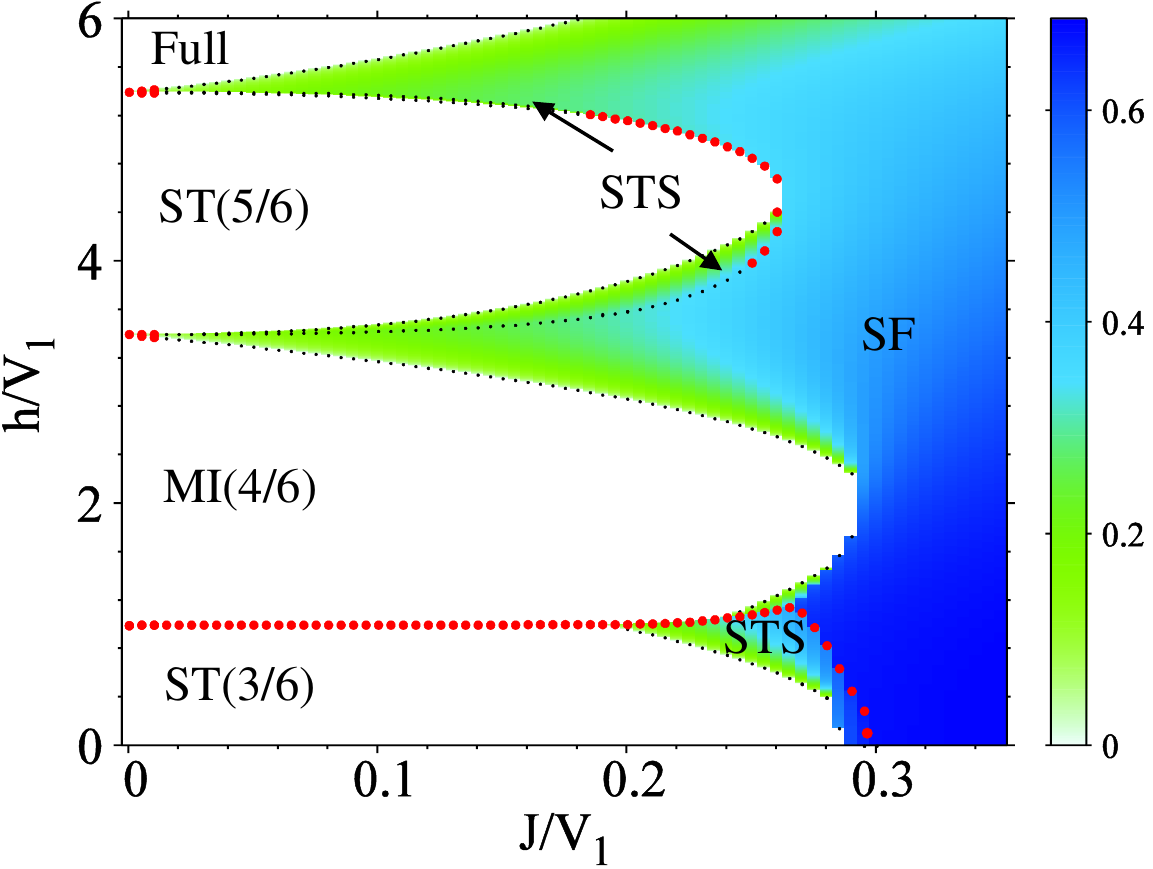}
\includegraphics[width=77mm]{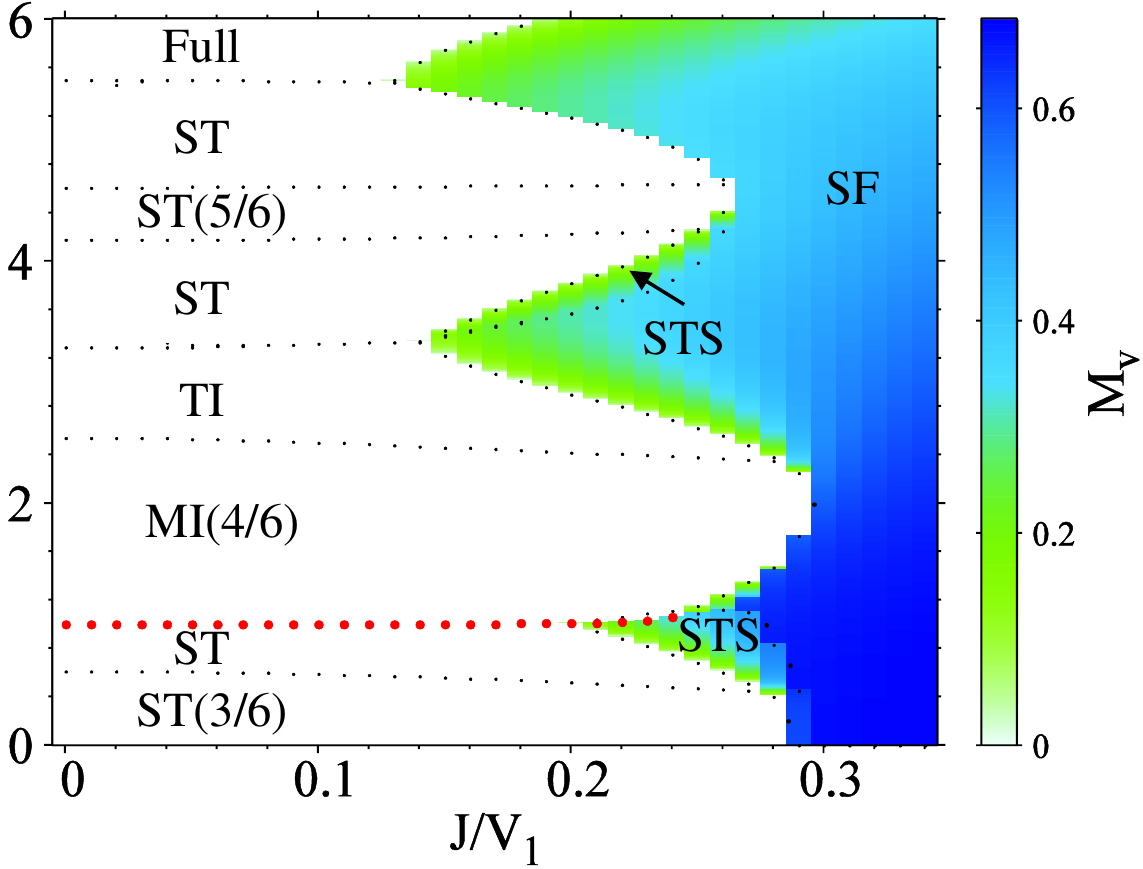}
\caption{(Color online) Left: ground state $J-h$ phase diagram of the
CAS model for $\frac{V_2}{V_1}=0.6$. Right: $J-h$ phase diagram of
the CAS model for $\frac{V_2}{V_1}=0.6$ at temperature $T/V_1=0.1$.
The same as the SMS model, the density of colors shows amount of off
diagonal order parameter, and the red (black) dotted lines show
first-order (second-order) phase transitions. The different orders
are defined in the table \ref{tab:CAS}. }\label{fig:CAS-Jh-atT}
\end{figure}
%%%%%%%%%%%%%%%%%%%%%%%%%%%%%%%%%%%%%%%%%%%%%%%%
%%%%%%%%%%%%%%%%%%%%%%%%%%%%%%%%%%%%%%%%%%%%% Definition of the orders  %%%%%%%%%%%%%%%%%%%%%%%%%%%%%%%%%%%%%%%%%%%%%
\begin{table}
\caption{Definitions of various ground state phases of the CAS model. Because of
the NNN interactions we divide the subsystem with spin $\s=1/2$
($\t=1$) into two sublattices A and B (C and
D) \cite{heydarinasab2018spin}. The longitudinal magnetizations
$m_{A(B)}^z=\la \s_{A(B)}^z \ra$ and $M_{C(D)}^z=\la \t_{C(D)}^z
\ra$, and also the total transverse magnetization $M_v$ are obtained
using cluster mean field theory.} \label{tab:CAS}
%\begin{ruledtabular}
\begin{center}
\begin{tabular}{ccc}
\br
Phases & sublattices magnetizations & $M_v$\\
\hline
SF& $m_A^z=m_B^z$, ~$M_C^z=M_D^z$& $\neq 0$\\
Full& $m_A^z=m_B^z=1/2$, ~$M_C^z=M_D^z=1$& 0\\
MI(4/6)& $m_A^z=m_B^z$, ~$M_C^z=M_D^z$& 0\\
ST(3/6) & $m_A^z \neq m_B^z$,~ $M_C^z \neq M_D^z$& 0\\
ST(5/6) & $m_A^z \neq m_B^z$,~ $M_C^z \neq M_D^z$& 0\\
STS & $m_A^z \neq m_B^z$,~ $M_C^z \neq M_D^z$& $\neq 0$\\
\br
\end{tabular}
\end{center}
%\end{ruledtabular}
\end{table}
%%%%%%%%%%%%%%%%%%%%%%%%%%%%%%
The ground state phase diagram of the CAS model for the frustration
$\frac{V_2}{V_1}=0.6$ is plotted in the left panel of Fig.
\ref{fig:CAS-Jh-atT}. Various ground state phases of this model is defined in the table \ref{tab:CAS}. 

\begin{figure}[h]
\centerline{\includegraphics[width=80mm]{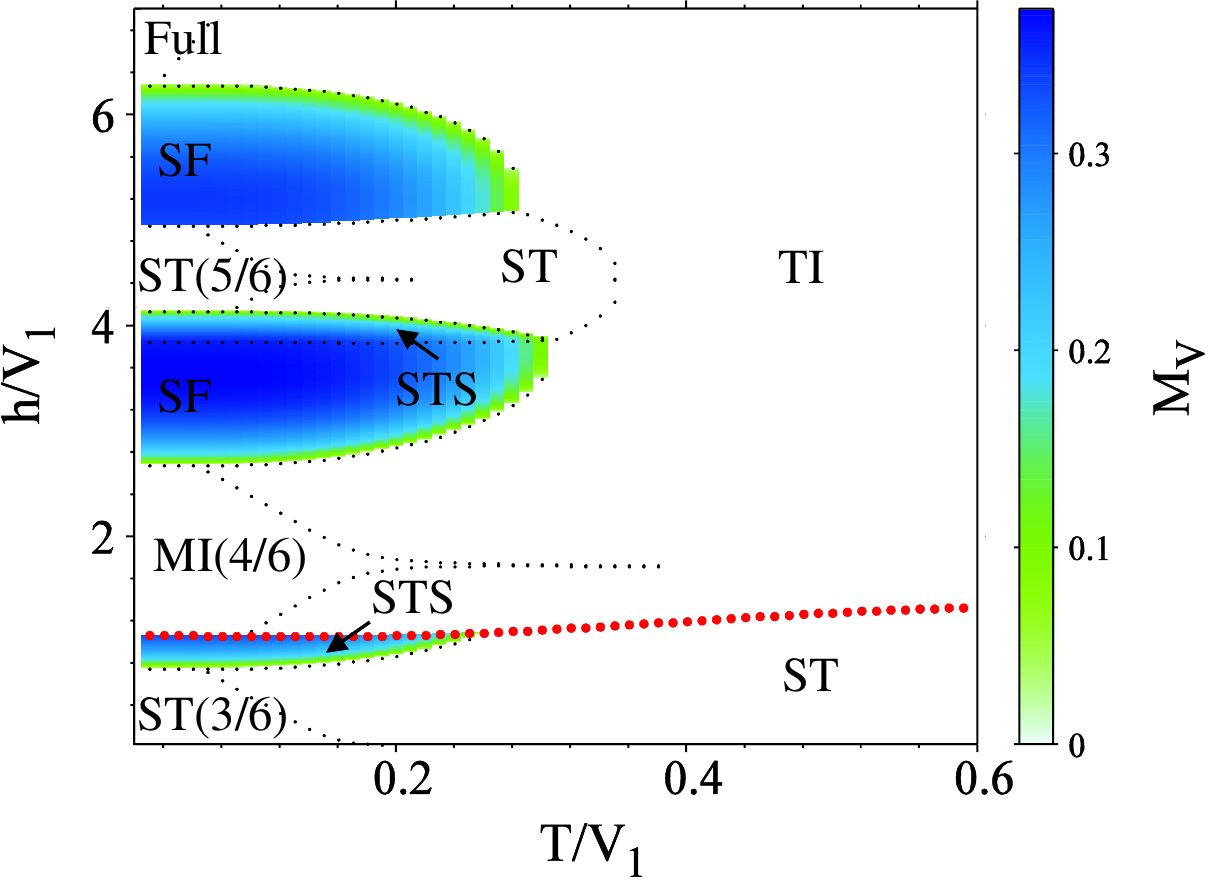}}
\caption{(Color online) $T-h$ phase diagram of the CAS model for
$\frac{V_2}{V_1}=0.6$ at the line $\frac{J}{V_1}=0.24$, where all
phases exist in the ground state phase diagram. The red (black)
dotted lines show first-order (second-order) phase transitions.}
\label{fig:LPD-Th}
\end{figure}

For small values of $J/V_1$, the translational
symmetry of both subsystems breaks for weak and strong magnetic
fields and stripe solids ST(3/6) and ST(5/6) with different filling
factors appear in the ground state phase diagram. At moderate
magnetic fields, the translational symmetries of both subsystems are
however restored in the MI(4/6) Mott insulating phase, where the
spins 1 (spins 1/2) align parallel (anti-parallel) to the magnetic
field. Strong enough magnetic fields ($h/V_1\gtrsim 5.5$) align all
spins and the system enters to the Mott insulating Full phase. An
enhancement of the hopping energy breaks the U(1) symmetry, and adds
the superfluid component to the stripe solids and Mott insulating
phases. Therefore stripe supersolid (STS) and superfluid (SF) appear
around the stripe solids and Mott insulator, respectively. While stripe solids to STS and SF transitions, and MI(4/6)-SF phase transitions are of second order, MI(4/6)-STS transition is of first order for all parameters ranges. This transition remains first order even when these phases vanishes at finite temperature. Therefore TI-STS and TI-ST transitions are first order in all ranges of $T$ (see Fig. \ref{fig:LPD-Th}).

For $\frac{V_2}{V_1}<0.5$, the translational symmetry breaks in the
presence of a moderate magnetic field and the ST(4/6) solid emerges
instead of the MI(4/6) Mott insulator (not shown).

In the presence of thermal fluctuations, similar to the SMS model,
magnetization plateaus melt and a thermal stripe solid
(ST) appears around the ST(3/6) and ST(5/6) phases, as shown in the
right panel of Fig. \ref{fig:CAS-Jh-atT}. In this phase the magnetization varies with the magnetic field, and the translational symmetry of both subsystems breaks.

Also melting process results in the emergence of a thermal insulator around the MI(4/6) and Full phases, where both the U(1) symmetry and the translational symmetry of the CAS lattice are preserved, however the magnetization varies with $h$. Moreover thermal
fluctuations destroy the superfluid order and causes the STS and SF
phases to be disappeared for $J/V_1$ around $T/V_1$.
%%%%%%%%%%%%%%%%%%%%%%%%%% FIG 0 %%%%%%%%%%%%%%%%%%
%%%%%%%%%%%%%%%%%%%%%%%%%%%%%%%%%%%%%%%%%%%%%%%%

We have also plotted in Fig. \ref{fig:LPD-Th}, the $T-h$ phase
diagram of the CAS model for $\frac{V_2}{V_1}=0.6$ and $\frac{J}{V_1}=0.24$ to find the
transition temperature for different phases. 
Thermal fluctuations melt ST(3/6) and ST(4/6) solids, and the MI(4/6) Mott insulator and transform them respectively to the thermal solid and thermal insulator. By further
increasing of temperature the ST phase also transform to the TI
phase where both translational symmetries restore but magnetization does not
vanish. Moreover thermal fluctuations suppress superfluid component
and cause the SF-TI transition. However, there are a narrow region
at larger magnetic fields in which increasing temperature breaks
translational symmetry instead, and SF-ST transition occurs. Further increasing temperature returns translational symmetry at ST-TI transition. 
Increasing temperature in all ranges of supersolid phase restores
the U(1) symmetry by STS-ST transition. 
As this figure shows, at finite temperature, the STS-MI(4/6) transition remains first order, even when these orders wash out completely the ST-TI transition remains first order. 

Similar to SMS model the transition temperature for the solid phases is larger far from the borders. The SF-TI critical temperature strongly depends on the values of hopping energy, so that by increasing the hopping energy the SF-TI critical temperature increases. For $\frac{J}{V_1}=0.21$ this temperature is $\frac{T}{V_1}=0.25$ and for $\frac{J}{V_1}=0.24$ it is $\frac{T}{V_1}=0.28$. Since the superfluid order is more fragile than solid order, the supersolid transition temperature is controlled by hopping energy. STS phase which is formed at lower magnetic field transforms to ST solid phase at temperature $\frac{T}{V_1}=0.14$ in $\frac{J}{V_1}=0.21$, while this transition happens at $\frac{T}{V_1}=0.27$ in larger hopping energy $\frac{J}{V_1}=0.24$. Also the STS phase which is formed at larger magnetic field persist up to temperature $\frac{T}{V_1}=0.25$ for $\frac{J}{V_1}=0.21$, while this phase is present up to temperature $\frac{T}{V_1}=0.3$ at $\frac{J}{V_1}=0.24$.

%%%%%%%%%%%% Thermodynamic functions %%%%%%%%%%%%%%%%%%
\section {Thermodynamic functions and magnetocaloric effect} \label{Thermodynamic-functions}
%%%%%%%%%%%%%%%%%%%%%%%%%%%%%%%%%%%%%%%%%%%%%%%%%%%%%%%

In this section we investigate the behavior of various thermodynamic
functions as well as the magnetocaloric effect in the mixed
spin-(1,1/2) model with the SMS and CAS arrangements. Using CMFT
(see the appendix), we have obtained the magnetization, entropy,
specific heat and also investigated the behavior of the
magnetocaloric effect. Magnetic susceptibility and specific heat
respectively demonstrate the amount of thermal fluctuations in the
magnetization and internal energy, while magnetocaloric effect
contains both of these fluctuations.

%%%%%%%%%%%%%%%%%%%%%%%%%%%%%%%%%%%%%%%%%%%%%%%%%%%%%%%%%%%%%%%%%%%%%%%%%%%%%%%%%%%%
\subsection{Isothermal variations of thermodynamic functions}
In this subsection, we investigate the isothermal variations of
mentioned thermodynamic functions in different solids, supersolids,
Mott insulators and superfluid phases.

\subsubsection{Magnetization}\label{sec:intro-kniz}
%%%%%%%%%%%%%%%%%%%%%%%%%%%%%%%%%%%%%%%%%%
\begin{figure*}[ht!]
\centering
\includegraphics[width=79mm]{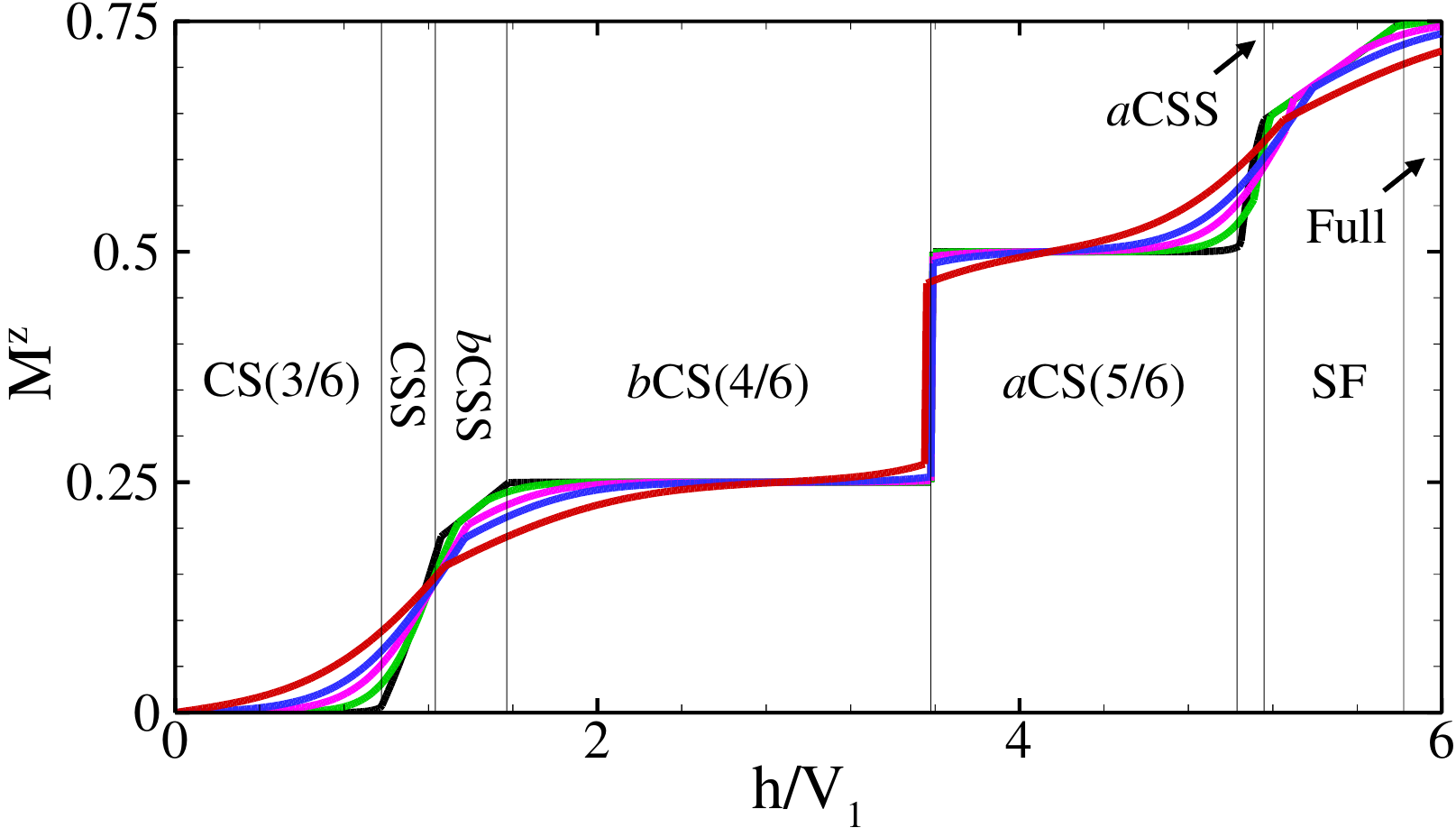}
\includegraphics[width=75mm]{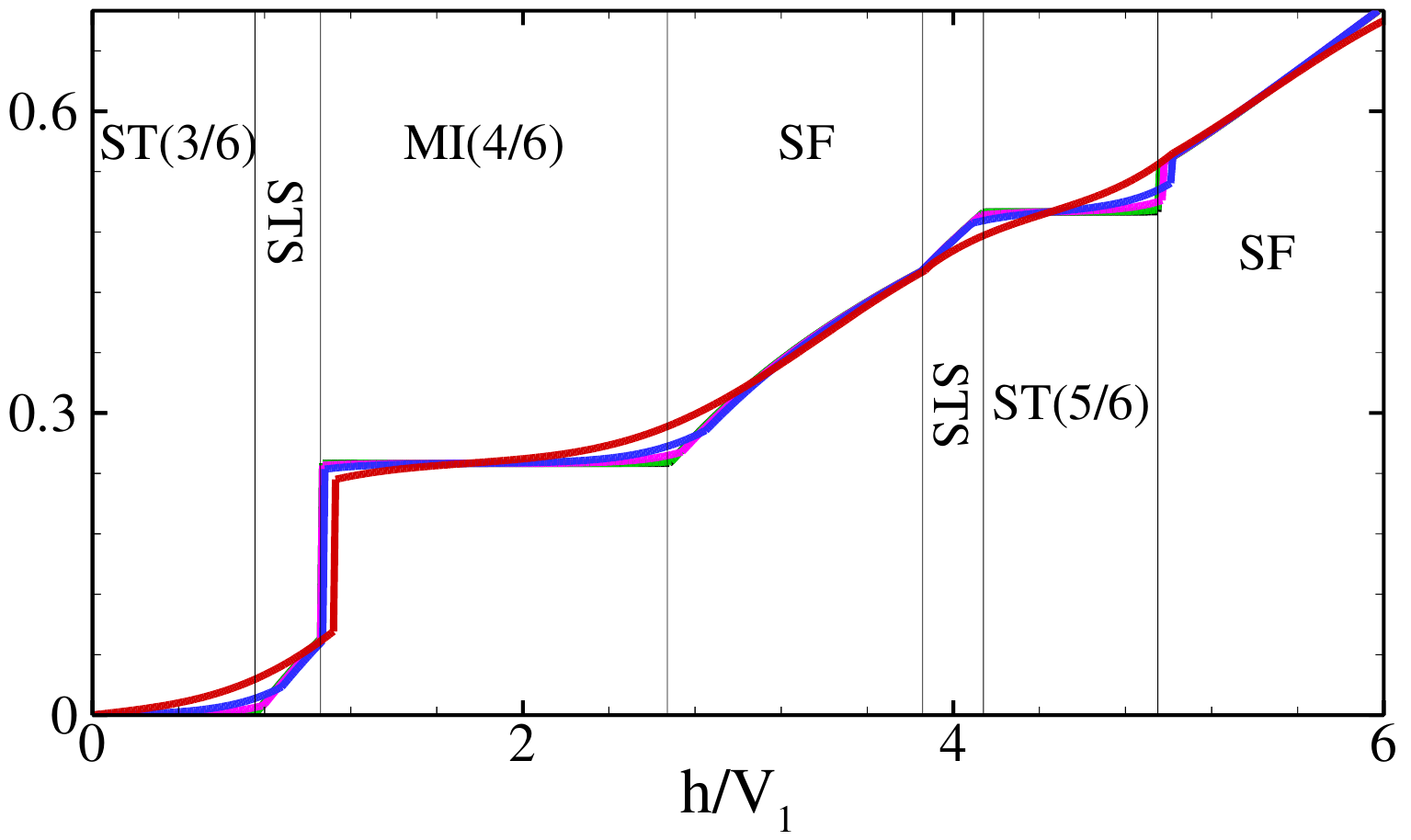}
\caption{(Color online) Longitudinal magnetization versus magnetic
field $h/V_1$, at $\frac{V_2}{V_1}=0.6$, for different temperatures $T/V_1$. Black: 0.05, green: 0.1, magenta: 0.15, blue: 0.2 and red: 0.3.
Left: for SMS model at $J/V_1=0.165$, and right: for CAS model at
$J/V_1=0.24$.} \label{fig:MZh-T}
\end{figure*}
%%%%%%%%%%%%%%%%%%%%%%%%%%%%%%%%%%%%%%%%%%
The longitudinal magnetization is obtained from the following relation:
\begin{eqnarray}
\label{eq:CMF-MZ-Xi} m(T,h) &=& \frac12 (\langle \s^z \rangle +
\langle \t^z \rangle),
\end{eqnarray}
where $\la\dots\ra$ denotes the ensemble average, computing from Eq.
(\ref{eq:CMFT-ave}). The
magnetization versus magnetic field is
plotted in Fig. \ref{fig:MZh-T} for different temperatures. The vertical lines in these figures show the ground state phase borders.

In the absence of temperature, for both the SMS and CAS models, the longitudinal magnetization increases
by increasing the magnetic field, passes through three nontrivial mid-plateaus
$m=0$, $m=0.25$ and $m=0.5$, and finally saturate at large saturation field (roughly around $h/V_1=6$), as
shown in Fig. \ref{fig:MZh-T}. These mid-plateau states correspond
to the mentioned solid or Mott insulator phases in the Figs. \ref{fig:SMS-Jh-atT} and \ref{fig:CAS-Jh-atT}, where longitudinal susceptibility
is zero. Due to stronger quantum fluctuations at finite temperature, the magnetization curve softens around
these phases' borders, where thermal solid and thermal insulator appear
and susceptibility increases. Mid-plateaus' widths depend on
temperature, they become smaller by increasing temperature and
vanish at a critical temperature, as seen in Figs. \ref{fig:CPD-Th} and \ref{fig:LPD-Th}.

At low temperatures, any break in the plots of the magnetization and susceptibility versus $h/V_1$, indicates the second order
phase transition. However, at the first order transition points, the
magnetization shows a jump at the critical field and the
susceptibility diverges. The $b$CS(4/6)-$a$CS(5/6) transition in the
SMS model and the STS-MI(4/6) transition in the CAS model are of first
order, while all other transitions are of second order. Thermal
fluctuations soften the transitions, by emerging thermal
solids and thermal insulator phases around the ground state solids
and Mott insulators, however for the $b$CS(4/6)-$a$CS(5/6)
transition at $h/V_1 \simeq 3.6$ in the SMS model and
the STS-MI(4/6) transition at $h/V_1 \simeq 1.1$ in the CAS model,
thermal fluctuations are not able to destroy the discontinuity in the
magnetization and these transitions remain first order (see
Fig. \ref{fig:MZh-T}). These jumps in the magnetization plot survive even though the plateaus around the discontinuity melt completely at higher
temperatures. This means that
the $b$CS-$a$CS and the $b$CS-TI transitions in the SMS model, and
the STS-TI and ST-TI transitions in the CAS model are always first order
(see also Figs. \ref{fig:CPD-Th} and \ref{fig:LPD-Th}).

%%%%%%%%%%%%%%%%%%%%%%%%%%%%%%%%%%%%%%%%%%%%%%%%%%%%%%%%%%%%%%%%%%%%%%%%%%%%%%%%%%%%%%%%%%%%%%%%%%%%%%%%%%%%%%%%%%%%%%%%%%
\subsubsection{Entropy and specific heat}\label{sec:intro-S-CV}
%%%%%%%%%%%%%%%%%%%%%%%%%%%%%%%%%%%%%%%%%%%%%%%%%%%%%%%%%%%%%%%%%%%%%%%%%%%%%%%%%%%%%%%%%%%%%%%%%%%%%%%%%%%%%%%%%%%%%%%%%%
\begin{figure*}[ht!]
\centering
\includegraphics[width=79mm]{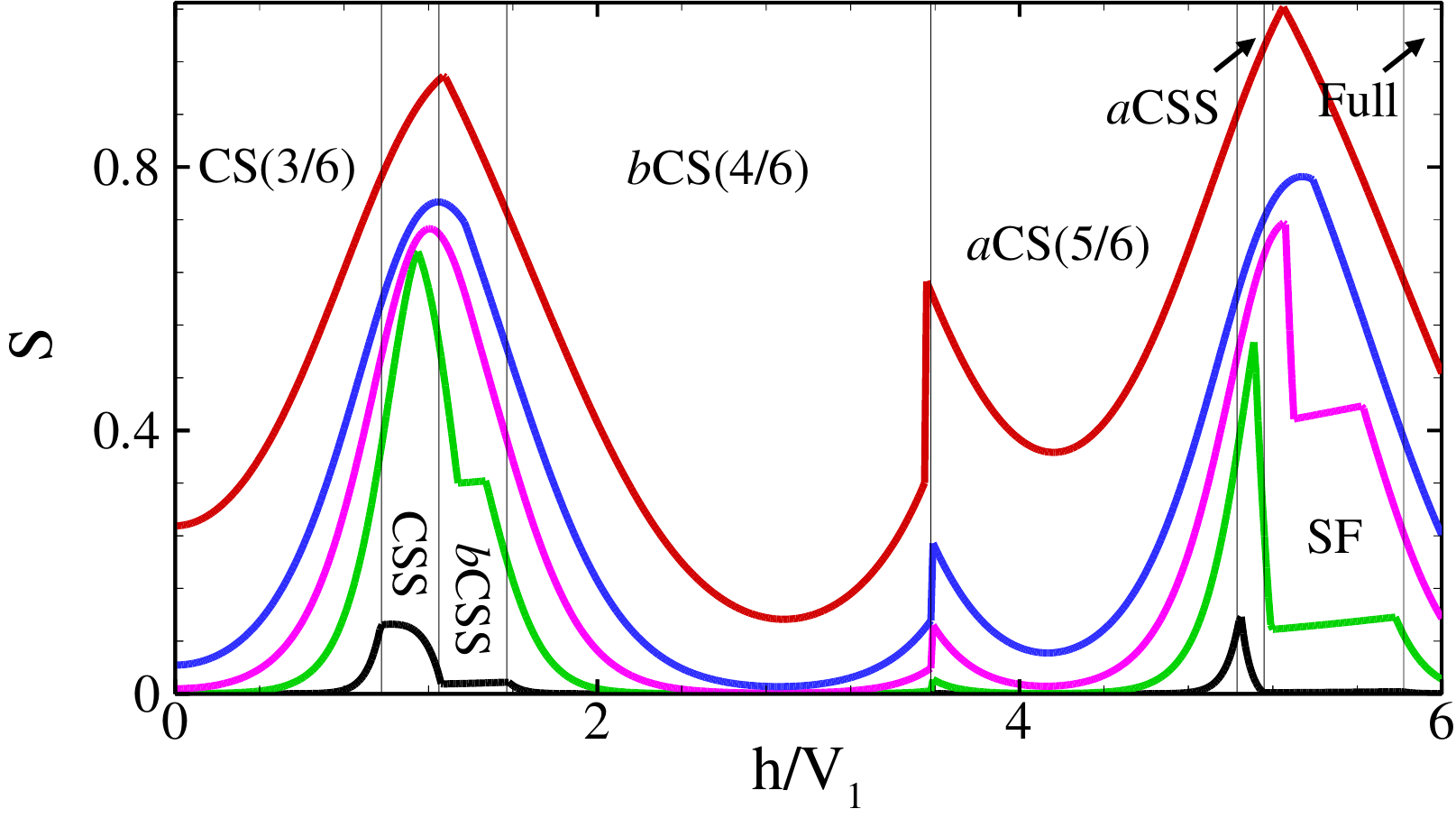}
\includegraphics[width=75mm]{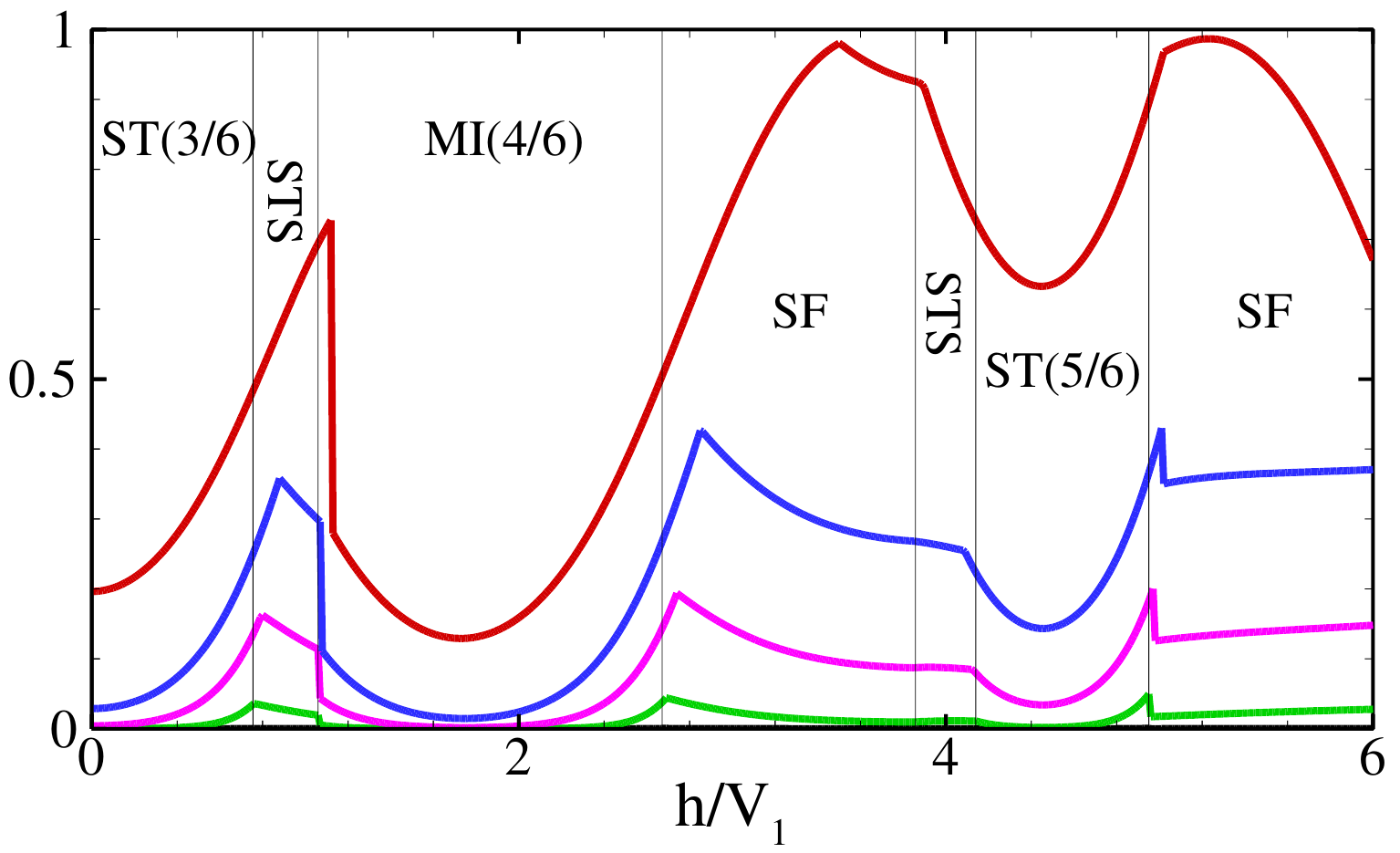}
\includegraphics[width=79mm]{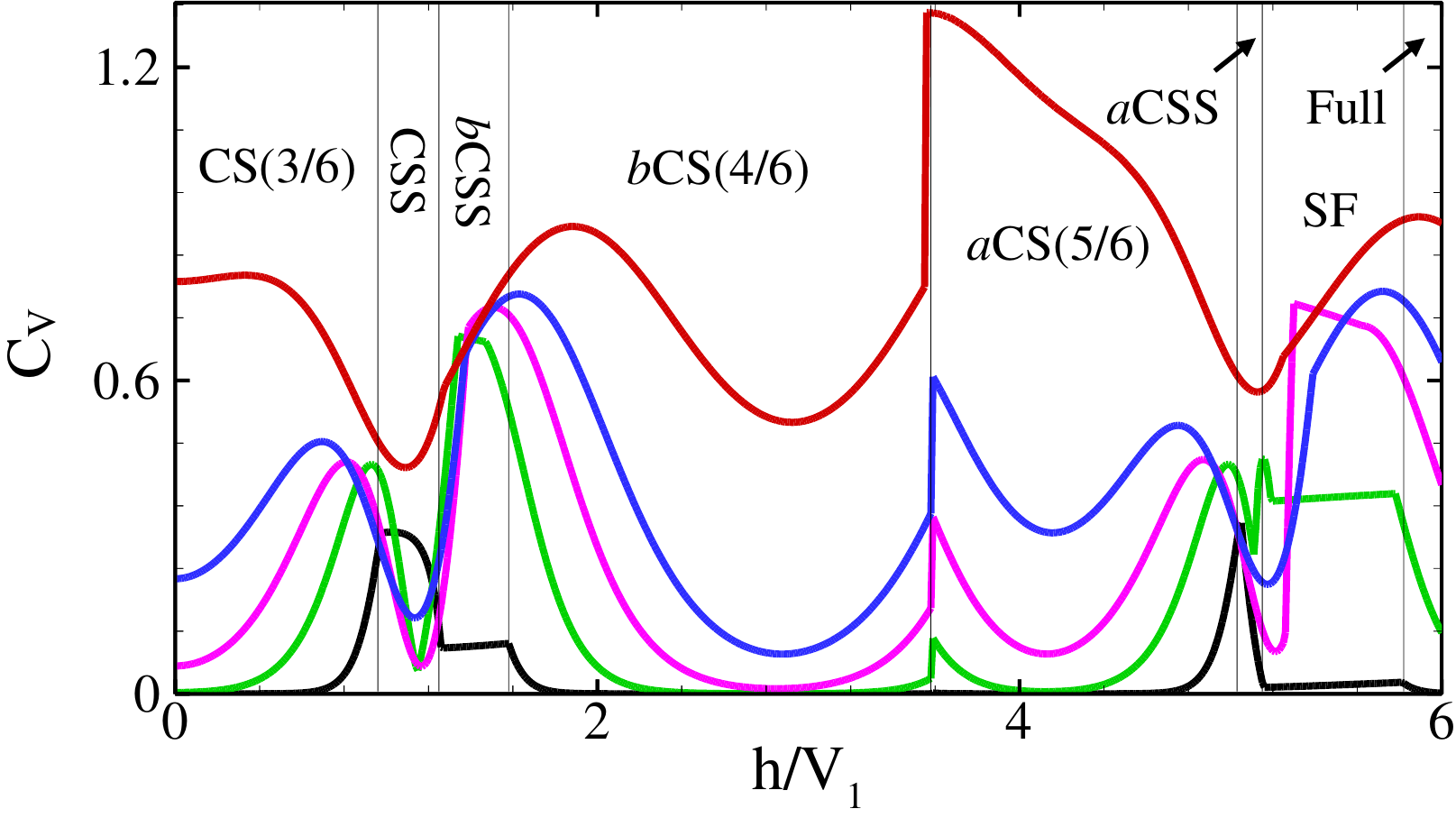}
\includegraphics[width=75mm]{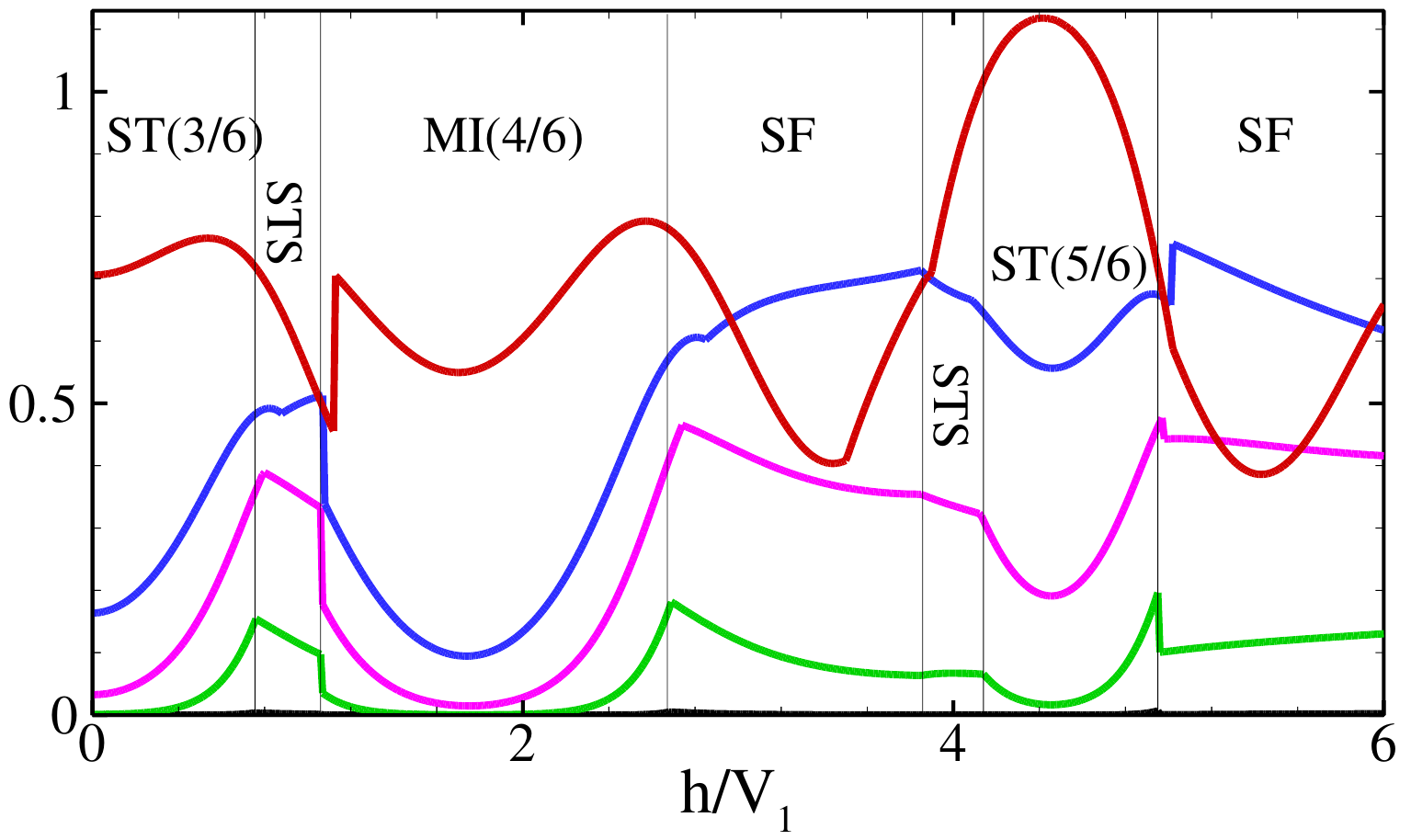}
\caption{(Color online) Entropy and specific heat versus magnetic
field $h/V_1$, at $\frac{V_2}{V_1}=0.6$, for different values of
temperature $T/V_1$. Black: 0.05, green: 0.1, magenta: 0.15, blue: 0.2 and red: 0.3. Left: for SMS model at $J/V_1=0.165$, and right: for
CAS model at $J/V_1=0.24$. } \label{fig:CVh-T}
\end{figure*}
%%%%%%%%%%%%%%%%%%%%%%%%%%%%%%%%%%%%%%%%%%
The entropy ($S$) and the specific heat ($C_V$) of the SMS and CAS
models, are obtained from the following relations:
%%%%%%%%%%%%%%%%%%
\begin{eqnarray}
\label{eq:CMF-S-CV}
\no S(T,h) &=& k_{\rm{B}} \ln\mathcal{Z}(T,h)+\frac1T \langle H(h) \rangle,\\
C_V(T,h) &=& \frac{1}{k_{\rm B}T^2} (\langle (H(h))^2 \rangle
-\langle H(h) \rangle^2),\label{eq:S-Cv}
\end{eqnarray}
%%%%%%%%%%%%%%%%%%
where averages are computed from Eq. (\ref{eq:CMFT-ave}) and $k_{\rm{B}}$ is the Boltzmann constant. We have plotted in the top panel of Fig. \ref{fig:CVh-T}, the entropy of the
SMS and CAS models versus $h/V_1$ for different temperatures.
At zero temperature the ensemble averages in Eq.
(\ref{eq:S-Cv}), reduce to the expectation values on the ground
state of the systems, and we obtain zero entropy for both
models. By increasing
temperature, higher energy eigenstates get occupied and the entropy
increases. At finite temperatures, the entropy strongly depends
on the magnetic field. It shows a peak or changes behavior at
critical fields, shows a jump at first order transition points and
is minimum in a plateau state. For example for the SMS model, at
$T/V_1=0.2$, where CSS and $b$CSS supersolids transform respectively to
the CS and $b$CS solids, the entropy increases gradually by increasing
$h$, passes through a maximum at the CS-$b$CS critical field and
becomes minimum in the $b$CS(4/6) plateau state. By more increasing
of the magnetic field, the entropy increases and suddenly jumps up
to a larger value at the first order $b$CS-$a$CS transition point. 
%Principally, entropy is large around the borders of solids and Mott insulator regions, so that entropy decreases around these phases with $h$ and after a constant regions in these plateaux phases, it increases with magnetic field at the larger $h$.
%However when the plateaus melt completely and these phases transform to the thermal solid and thermal insulator phases, entropy shows minimum instead of constant regions.
Also at low $T$, for example at $T/V_1=0.1$, for both models the entropy approximatly increases linearly with $h$ in the superfluid phases at larger magnetic fields. Also this is the case for the STS phase in the CAS model.

In order to see the effects of thermal fluctuations on the internal
energy of the SMS and CAS models we also investigate the behavior of the
specific heat. In the bottom panels of Fig. \ref{fig:CVh-T}, we have
plotted the specific heat $C_V$ versus $h/V_1$ for different
temperatures.
At zero temperature, both systems are in their ground state and the
specific heat is zero at all magnetic fields.
At a finite temperature, similar to the entropy, the
specific heat strongly depends on the magnetic field. It is constant deeply in the solid and Mott insulating phases, and increases around the thermal solid and thermal insulator phases. However at high
temperatures, for example at $T/V_1\simeq 0.3$, the specific heat shows a peak
in the thermal solids and thermal insulator. At low temperatures,
$C_V$ could be approximated by a linear function of $h$ in the
superfluid and supersolid phases, except for the CSS phase in the
SMS model which develops a peak and also for the SF phase at smaller
magnetic field in the CAS model.

%Moreover, at the first order transition fields the entropy and specific heat has a discontinuity and jumps up to a larger value, whiles there are breaks at the second order phase transitions.

%%%%%%%%%%%%%%%%%%%%%%%%%%%%%%%%%%%%%%%%%%%%%%%%%%%%%%%%%%%%%%%%%%%%%%%%%%%%%%%%%%%%%%%%%%%%%%%%%%%%%%%%%%%%%%
\subsubsection{Magnetocaloric effect}\label{sec:intro-MCE}
%%%%%%%%%%%%%%%%%%%%%%%%%%%%%%%%%%%%%%%%%%
\begin{figure*}[ht]
\centering
\includegraphics[width=79mm]{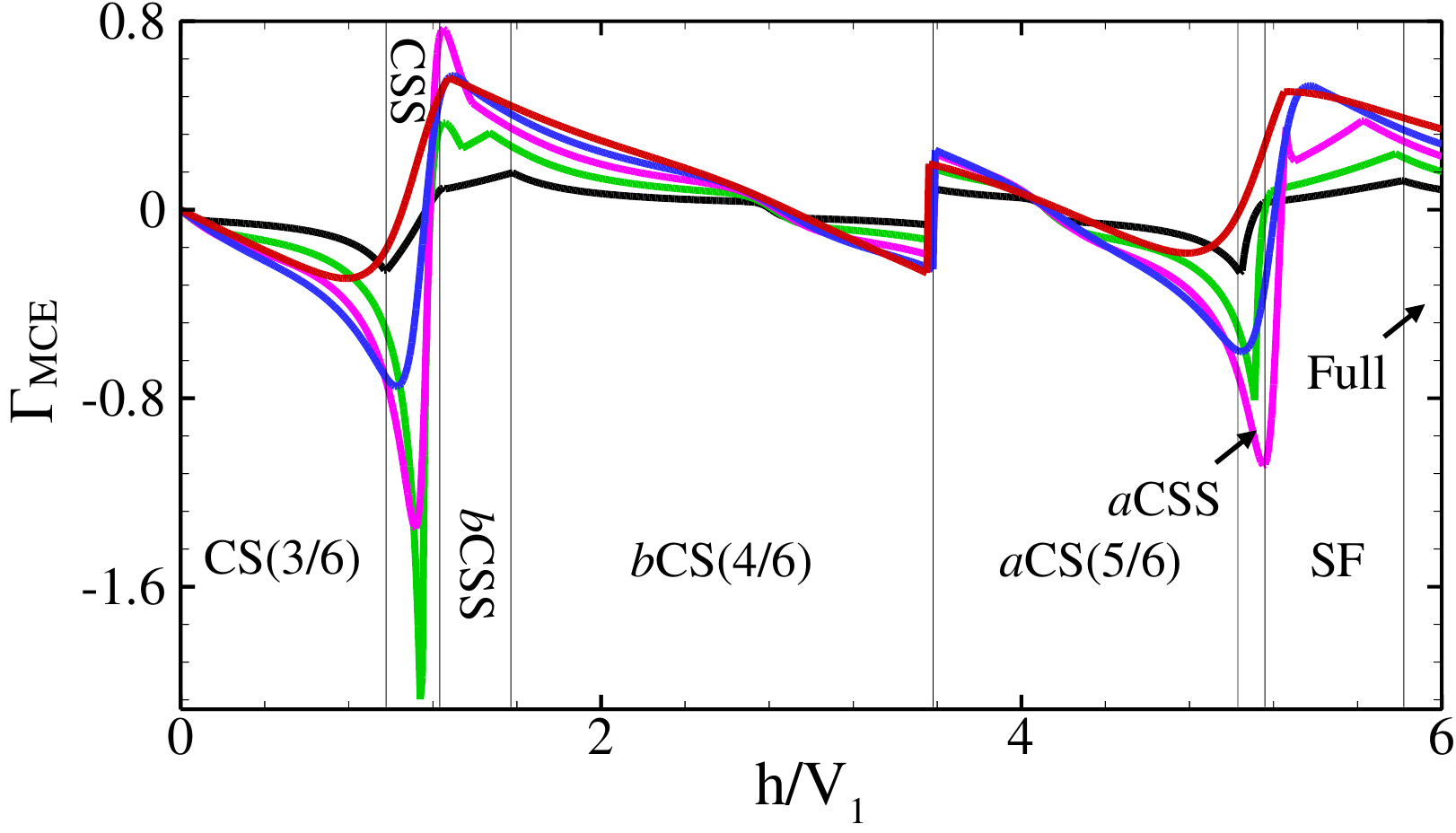}
\includegraphics[width=75mm]{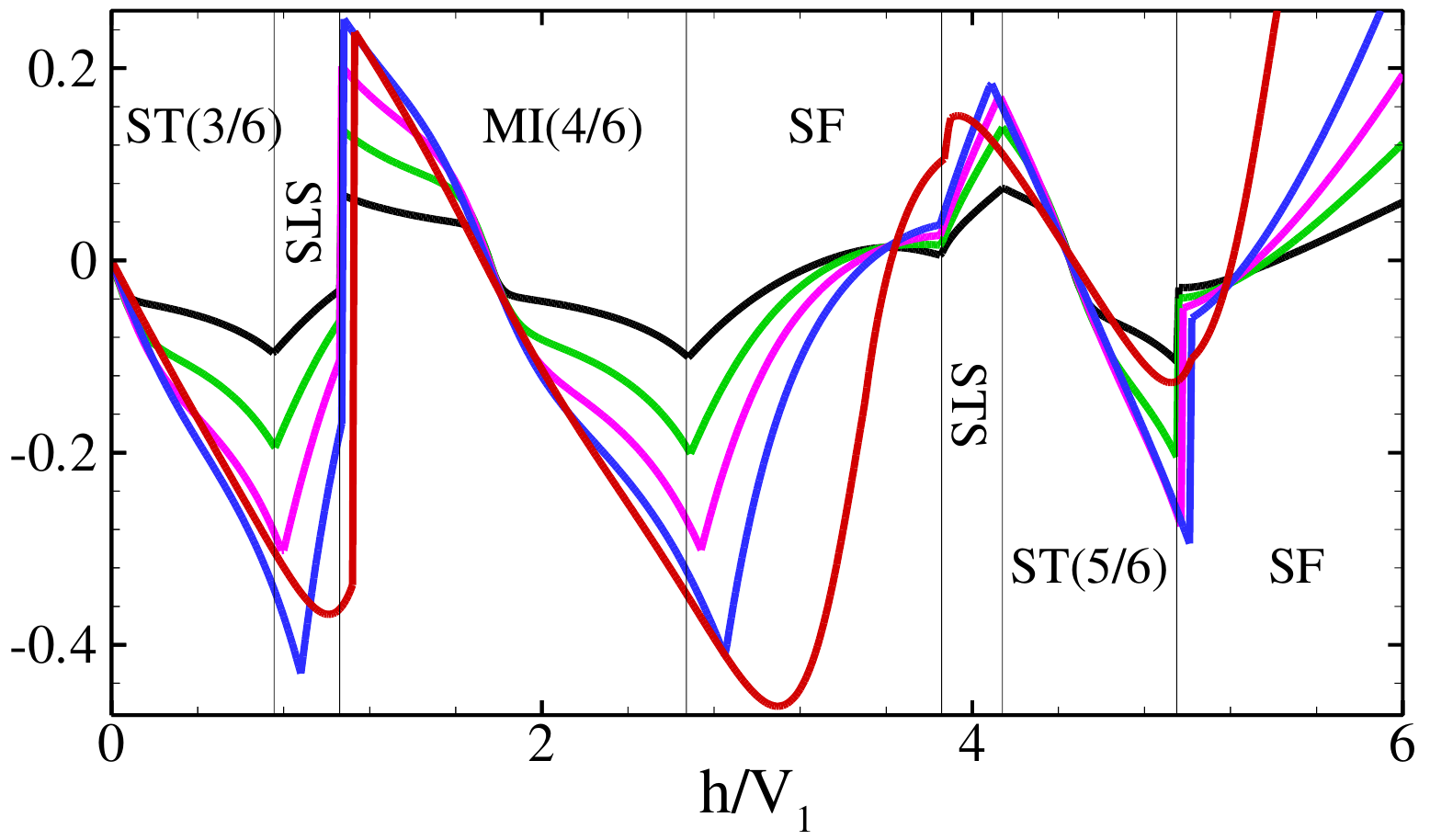}
\caption{(Color online) MCE versus magnetic field $h/V_1$, at
$\frac{V_2}{V_1}=0.6$, for different values of temperature $T/V_1$. Black: 0.05, green: 0.1, magenta: 0.15, blue: 0.2 and red: 0.3. Left:
for SMS model at $J/V_1=0.165$, and right: for CAS model at
$J/V_1=0.24$.} \label{fig:MCEh-T}
\end{figure*}
%%%%%%%%%%%%%%%%%%%%%%%%%%%%%%%%%%%%%%%%%%
%%%%%%%%%%%%%%%%%%%%%%%%%%%%%%%%%%%%%%%%%%
\begin{figure*}[ht!]
\centering
\includegraphics[width=75mm]{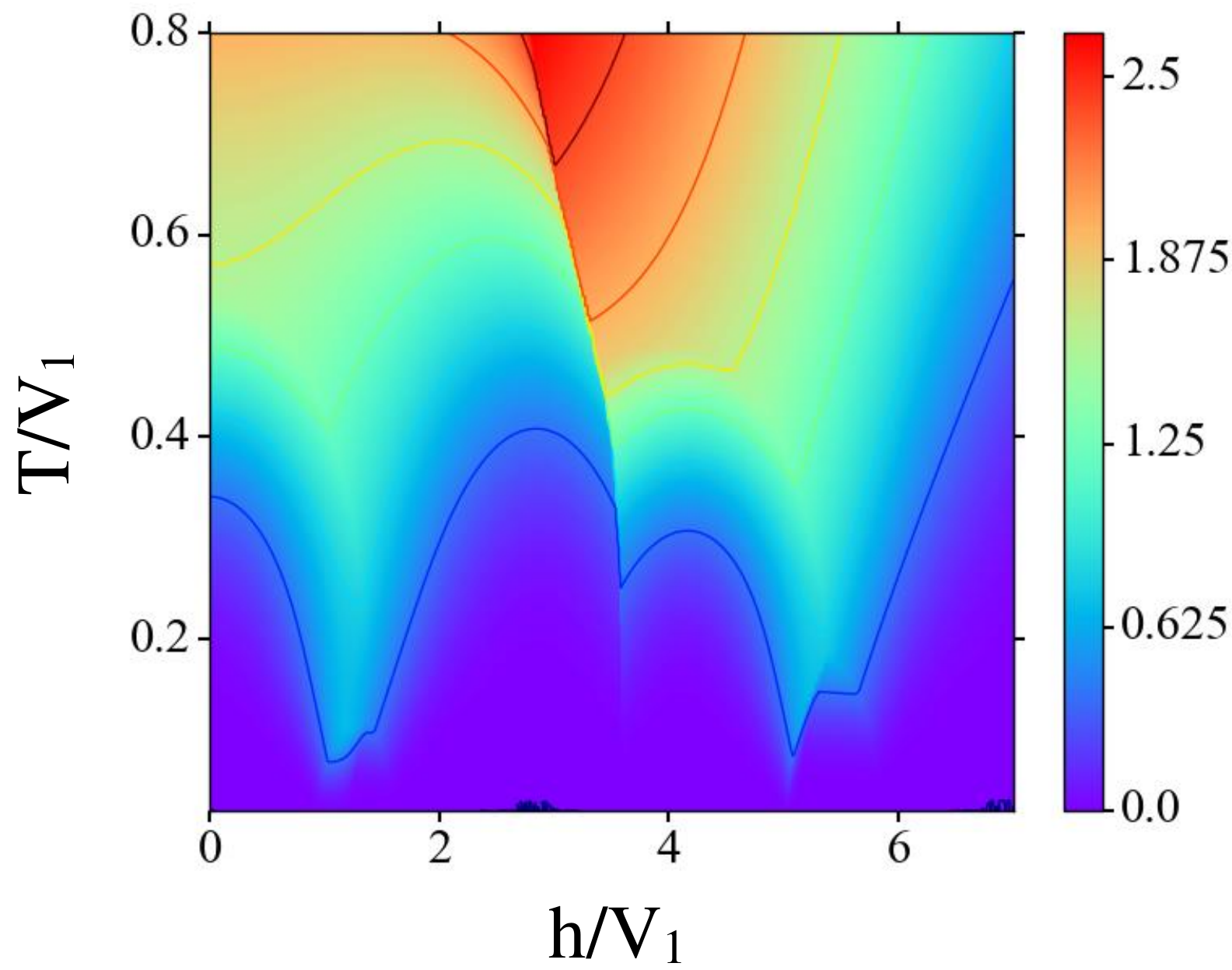}
\includegraphics[width=75mm]{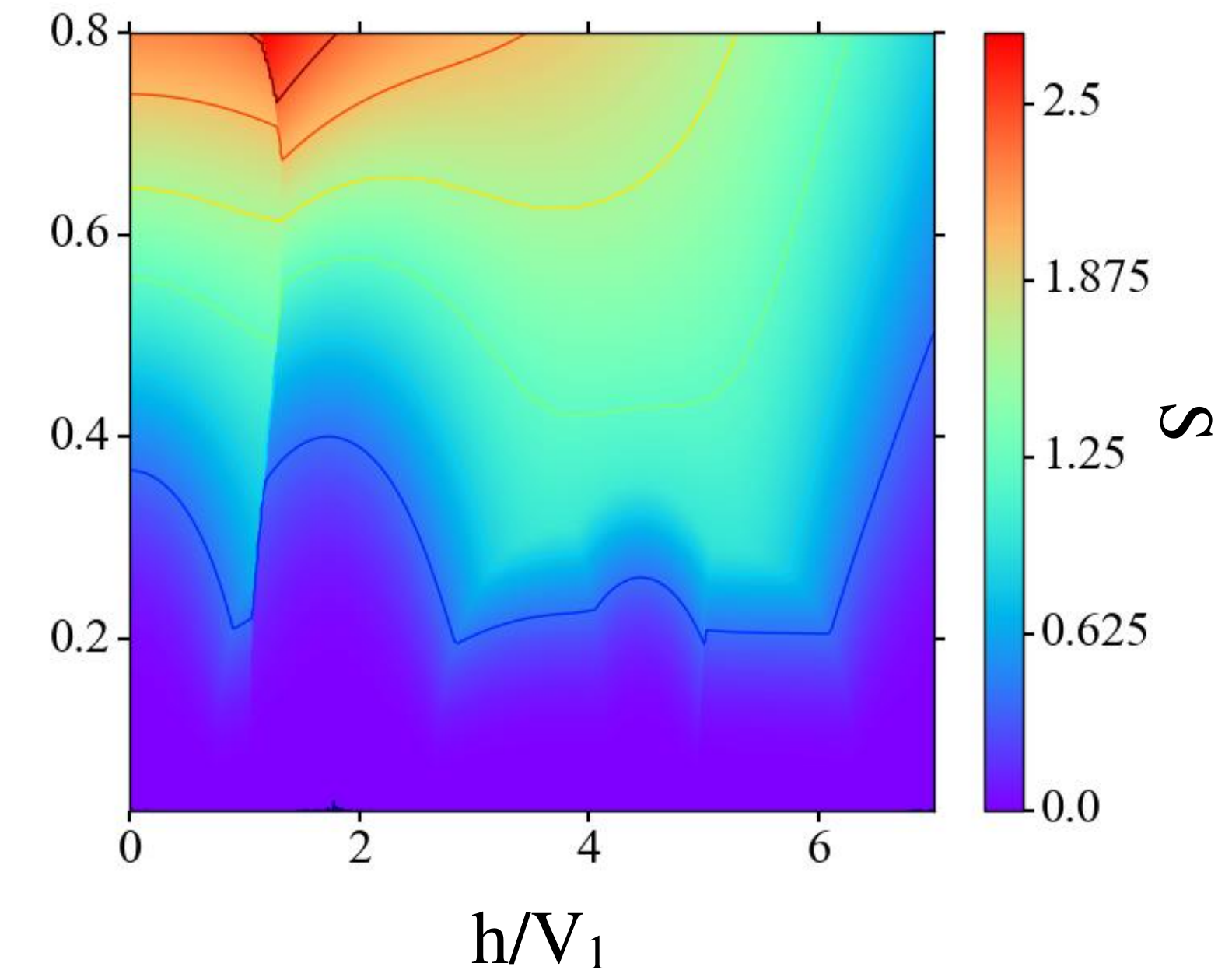}
\caption{(Color online) Isentrope  density plots at
$\frac{V_2}{V_1}=0.6$. Left: for SMS model at $J/V_1=0.165$, and
right: for CAS model at $J/V_1=0.24$.} \label{fig:MCE-S-h-T}
\end{figure*}
%%%%%%%%%%%%%%%%%%%%%%%%%%%%%%%%%%%%%%%%%%
MCE is measured by Gruneisen Parameter $\Gamma_{\rm{MCE}}$, defined as:
%%%%%%%%%%%%%%%%%%
\begin{eqnarray}
\label{eq:MCE1} \Gamma_{\rm{MCE}} = \left(\frac{\partial T}{\partial
h}\right)_S,
\end{eqnarray}
%%%%%%%%%%%%%%%%%%
where $h$, $T$ and $S$ are the magnetic field, temperature
and entropy of the system, respectively. Using cyclic relations
between these parameters, $\Gamma_{\rm{MCE}}$ is given in terms of the
specific heat and the magnetization of the systems as:
%%%%%%%%%%%%%%%%%%
\begin{eqnarray}
\label{eq:MCE2} \Gamma_{\rm{MCE}} = - \frac{(\frac{\partial
S}{\partial h})_T}{(\frac{\partial S}{\partial T})_h} = -
\frac{T}{C_V} \left(\frac{\partial m}{\partial T}\right)_h.
\end{eqnarray}
%%%%%%%%%%%%%%%%%%
In order to obtain $\Gamma_{\rm{MCE}}$ numerically, it is helpful to
simplify the above equation as:
\begin{eqnarray}
\label{eq:MCE3} \Gamma_{\rm{MCE}} = - T \frac{\langle H S^z \rangle
- \langle H \rangle \langle S^z \rangle}{\langle H^2 \rangle -
\langle H \rangle ^2}.
\end{eqnarray}
This relation indicates that fluctuations in both the magnetization and
the internal energy play essential role on the behavior of
$\Gamma_{\rm{MCE}}$. We have plotted in Fig. \ref{fig:MCEh-T}, the
parameter $\Gamma_{\rm{MCE}}$ versus magnetic field for different temperatures.

In the solid and the Mott insulating phases,
$\Gamma_{\rm{MCE}}$ changes sign and becomes negative at larger $h$.
This behavior which is a characteristic of ordered
phases \cite{schmidt2007magnetocaloric, garst2005sign}, could be
obtained from the magnetization. According to Eq. (\ref{eq:MCE2}),
$\Gamma_{\rm{MCE}}$ is proportional to $(\frac{\partial m}{\partial
T})_h$. In the presence of thermal fluctuations, the magnetization
changes inside the solid and Mott insulator phases, it decreases
(increases) by temperature at smaller (larger) $h$. This causes the
function $(\frac{\partial m}{\partial T})_h$ to be negative
(positive) around the solids and Mott insulators. Therefore
$\Gamma_{\rm{MCE}}$ changes sign in the thermal solid and thermal insulator phases. Moreover, at low $T$, MCE increases almost linearly in the
superfluid and supersolid phases, except for the $a$CSS in the SMS
model and the SF phase in the CAS model. MCE always is positive inside
SF and TI phases, which means that in these phases magnetic field always heats up the system.

At low temperatures, MCE increases or changes behavior at the second order transition points, while it has a discontinuity at the first order transition points. At low temperatures, the maximum cooling rate occurs in the vicinity of the solid-supersolid, solid-superfluid and MI(4/6)-superfluid quantum critical points, where there are a large accumulation of the entropy, see top panel of Fig. \ref{fig:CVh-T}. This increasing was expected from the relation between the Gruneisen parameter and entropy in Eq. \ref{eq:MCE2}. However, at higher temperatures the maximum cooling rate happens inside the thermal solid and thermal insulator phases.

We have also plotted in Fig. \ref{fig:MCE-S-h-T}, the density plot of the entropy in the $h-T$ phase diagram. This diagram would be useful for experimentalist. Actually, at low temperatures, critical points correspond to the minimums of the isentropes in the $h-T$ diagram \cite{schmidt2007magnetocaloric}. Hence MCE anomalies may be useful to map out the $h-T$ phase diagrams which are not accessible otherwise \cite{schmidt2007magnetocaloric, garst2005sign}. 

It is seen that entropy is constant in the solid and MI(4/6) phases and increases in the thermal solid and thermal insulating phases around them. Therefore isentropes develope a peak in these ranges of the $T-h$ phase diagram which confirms the sign changes of the MCE around the solid and MI(4/6) phases. Moreover, the linear behavior of the isentropes in some ranges of the superfluid and supersolid phases confirms the linear behavior of the MCE in these phases. 

In Fig. \ref{fig:MCE-S-h-T}, the first order transitions specify by non-continues changes of the isentropes and the second order transitions specify by changes in the behaviors of the isentropes around the transition points. At low temperatures, the minimums of the isentropes are around the solid-supersolid, solid-superfluid and MI(4/6)-superfluid critical fields, which confirms a large cooling rate at these points. Moreover at low temperature, the large positive values of $\Gamma_{\rm{MCE}}$ could be seen in the tricritical points around the superfluid and thermal solid or thermal insulator phases, where isentropes feel breaks. However at larger $T$ these points are placed inside the thermal solid and thermal insulator phases.

%%%%%%%%%%%%%%%%%%%%%%%%%%%%%%%%%%%%%%%%%%%%%%%%%%%%%%%%%%%%%%%%%%%%%%%%%%%%%%%%%%%%%%%%%%%%%%%%%%%%%%%%%%%%%%
\subsection{Temperature variations of the thermodynamic functions and magnetocaloric effect}
%%%%%%%%%%%%%%%%%%%%%%%%%%%%%%%%%%%%%%%%%%%%%%%%%%%%%%%%%%%%%%%%%%%%%%%%%%%%%%%%%%%%%%%%%%%%%%%%%%%%%%%%%%%%%%
%%%%%%%%%%%%%%%%%%%%%%%%%%%%%%%%%%%%%%%%%%%%%%%%%%
%%%%%%%%%%%%%%%%%%%%%%%%%% FIG 0 %%%%%%%%%%%%%%%%%%
\begin{figure*}[ht!]
\includegraphics[width=175mm]{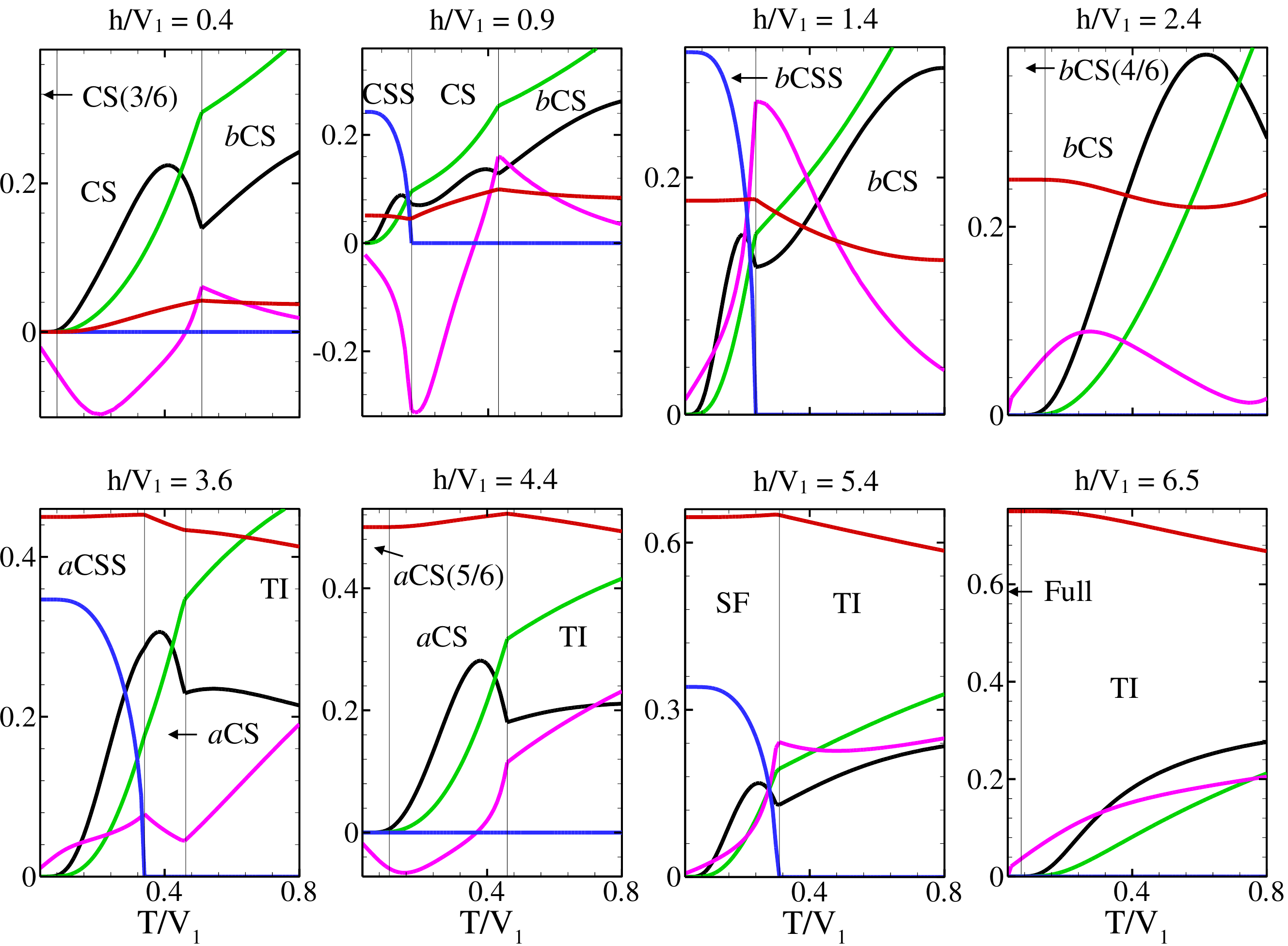}
\caption{(Color online) Different thermodynamic functions of SMS
model versus temperature $T/V_1$, for different values of $h/V_1$,
at $\frac{V_2}{V_1}=0.6$ and $\frac{J}{V_1}=0.22$. Red: the
longitudinal magnetization, blue: the transverse magnetization,
green: the scaled entropy $S/5$, black: the scaled specific heat
$C_V/5$ and magenta: the scaled MCE $\Gamma_{\rm{MCE}}/2$. In order to plot all thermodynamic functions in a single frame, we rescaled $S$, $C_V$ and $\Gamma_{\rm{MCE}}$.}
\label{fig:SMS-T}
\end{figure*}
%%%%%%%%%%%%%%%%%%%%%%%%%%%%%%%%%%%%%%%%%%%%%%%%

\begin{figure*}[ht!]
\includegraphics[width=175mm]{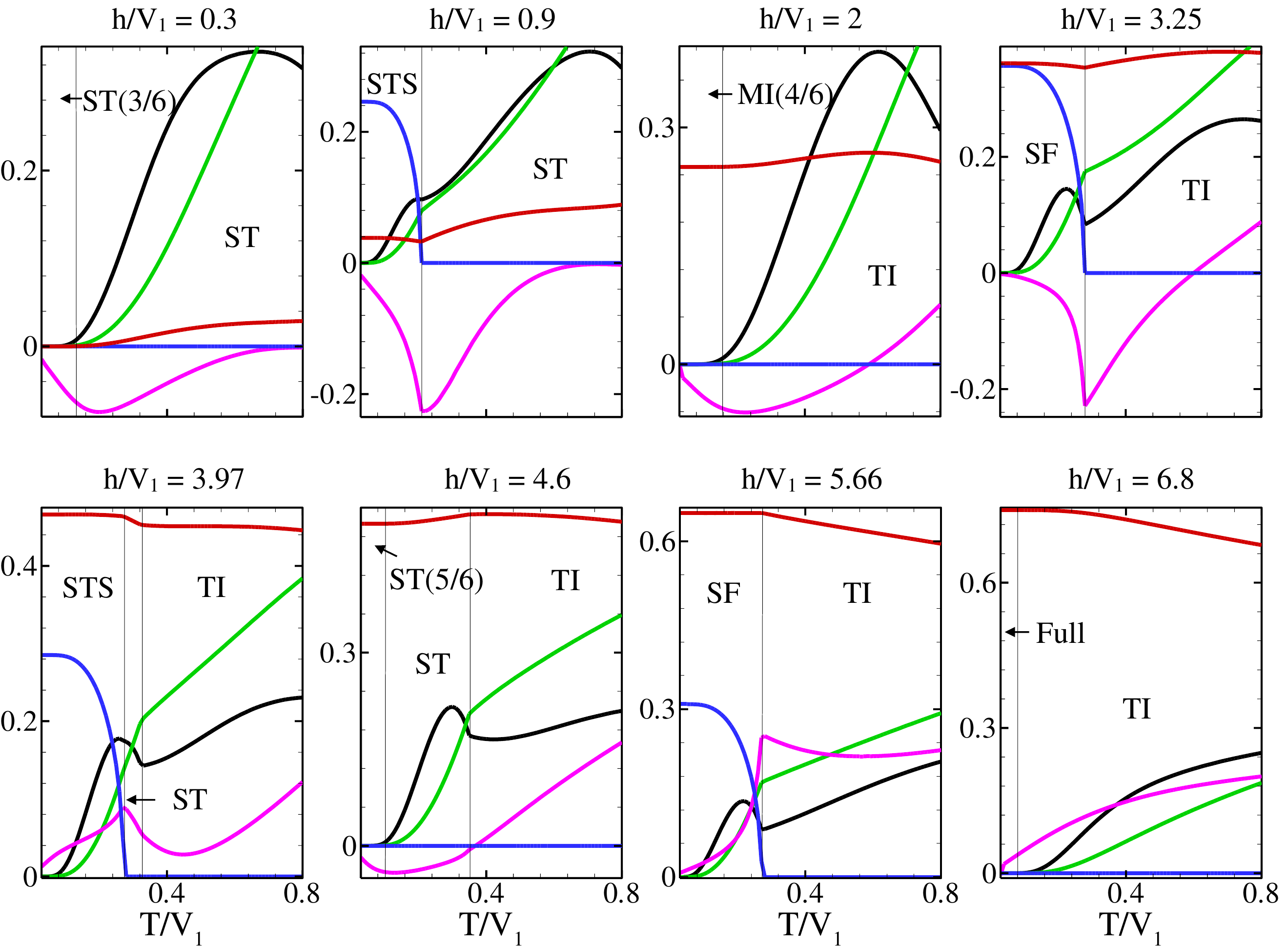}
\caption{(Color online) Different thermodynamic functions of CAS
model versus temperature $T/V_1$, for different values of $h/V_1$,
at $\frac{V_2}{V_1}=0.6$ and $\frac{J}{V_1}=0.24$. Red: the
longitudinal magnetization, blue: the transverse magnetization,
green: the scaled entropy $S/5$, black: the scaled specific heat
$C_V/5$ and magenta: the scaled MCE $\Gamma_{\rm{MCE}}/2$.}
\label{fig:CAS-T}
\end{figure*}

We have also plotted in Figs. \ref{fig:SMS-T} and \ref{fig:CAS-T}, the thermodynamic functions versus temperature for different values of magnetic field. 
In CS(3/6), $b$CS(4/6) and $a$CS(5/6) (in SMS model), and ST(3/6) and ST(5/6) (in CAS model) solid phases, where the magnetization shows plateaus, the system is gapped and both the entropy and the specific heat are zero, however the MCE is finite. $\Gamma_{\rm{MCE}}$ is positive in $b$CS(4/6) solid, while it's negative in all other ones. By increasing temperature and melting the magnetization plateaus, respectively thermal solid phases CS, $b$CS, $a$CS, and ST thermal solids emerge, where magnetization increases by temperature.
%Although increasing temperature melts the solid phases, but the magnetization increases with $T$ in the thermal solid phases. This subject may be related to the presence of the magnetic field $h$. In fact thermal fluctuations shield the quantum fluctuations, which were the reason of different solid orders. Therefore, at finite $h$ the spins fillip in the field direction, that result increasing the magnetization. Moreover specific heat has a peak in the thermal solid phases and entropy increases with $T$. MCE feels sign changes in the CS and $a$CS phases, however it's positive in the $b$CS phase. Further increasing $T$, enters the system to the TI phase where $C_V$ has a broad peak and magnetization decreases with temperature. Also entropy increase with $T$ in this phase and MCE is positive.
In all these thermal solid phases, the entropy increases by temperature and the specific heat possesses a peak. MCE has sign changes with $T$ in CS and $a$CS solids, however it's positive in $b$CS solid. In the ST solid phase the MCE depends on the magnetic field, it is negative at small magnetic fields, while positive at larger fields.

%Further increasing temperaure, the system enters to the TI phase. In this phase, by increasing temperature the magnetization decreases, the entropy increases, the specific heat has a broad peak and the MCE is positive.

%While superfluidity decreases by increasing temperatre in the SF phase, the longitudinal magnetization experiences small increasing.
In the SF phase, the specific heat shows a peak, MCE is positive and increases with temperature. The superfluid component is vanishing by increasing temperature where TI phase emerges. 
In the supersolid phases CSS, $b$CSS and $a$CSS, and STS, the entropy increases by temperature and the specific heat shows a peak.
The MCE is negative in the CSS phase, while it is positive in the $b$CSS and $a$CSS phases where magnetic field is stronger. In the STS supersolid, the MCE is negative at small magnetic fields, whereas it's positive at larger ones. By increasing $T$ and reduction of the superfluidity, in the $b$CSS, $a$CSS and STS supersolids, there is an small enhancement in the longitudinal magnetization, but the magnetization decreases in the CSS supersolid with low $h$. According to the values of $\Gamma_{\rm{MCE}}$, in the SMS model the maximum cooling rate occurs at the CSS-CS transition points. In CAS model, a large positive value of the MCE is seen at the SF-TI border with larger $h$, however there is a large cooling rate at the SF-TI transition point in the smaller magnetic field.

Finally, in the Mott insulating MI(4/6) and Full phases with the longitudinal magnetization plateaus, the entropy and specific heat are vanishing. The MCE  is negative in MI(4/6), while it's positive in the Full phase. At larger $T$ where these phases transform to the TI phase, the specific heat has a broad peak, magnetization decreases, entropy increases and MCE is positive.
%%%%%%%%%%%%%%%%%%%%%%%%%%%%%%%%%%%%%%%%%%%%%%%%%%%%%%%%%%%%%%%%%%%%%%%%%%%%%%%%%%%%%%%%%%%%%%%%%%%%%%%%%%%%%%
%%%%%%%%%%%%%%%%%%%%%%%%%% FIG 0 %%%%%%%%%%%%%%%%%%
%%%%%%%%%%%%%%%%%%%%%%%%%%%%%%%%%%%%%%%%%%%%%%%

In conclusion, the entropy increases with temperature in all phases of the both models. Also specific heat shows a narrow peak in the superfluid, supersolid and thermal solid phases, and develops a broad peak in the TI phase. 

%%%%%%%%%%%%%%%%%%%%%%%%%%%%%%%%%%%%%%%%%%%%%%%
%%%%%%%%%%%%%%%%%%% Summary %%%%%%%%%%%%%%%%%%%

\section{Summary and conclusion}\label{sec:summary}

To summarize, in the present paper, employing CMFT, we have studied thermal phase diagram of the mixed-spin (1,1/2) model on the square lattice with two different arrangements. We have demonstrated that at a finite temperature, thermal fluctuations soften the transitions, and different thermal solid and thermal insulators phases emerge around the ground state phases. 
Our results show that the solid-solid and supersolid-Mott insulator phase transitions maintain first order even at high temperatures where the ground state phases around these transition points are washed out completely. 
As the supersolid phase persists up to comparable temperature with the interaction terms, this model would be a playground for finding different supersolid phases in experiment.

In the second part of the paper we have also studied isothermal variations of different thermodynamic functions and magnetocaloric effect. Our results show that the MCE changes sign in the thermal solids and thermal insulator. At low temperatures, the maximum cooling rate is seen in the vicinity of the solid-supersolid, solid-superfluid and MI(4/6)-superfluid critical points, whereas at higher temperatures it placed inside the thermal solids and thermal insulator. At low temperatures, the large positive values of $\Gamma_{\rm{MCE}}$ was seen in the tricritical points between superfluid and thermal solid or TI phases, however they placed inside the thermal solid and TI phases at higher temperature. This work is the first study on the MCE in supersolid phases, and a large cooling rate around this phase in addition to the multi-peak structure of the specific heat could be a signature of these phases, which is useful for experimental detection of such phases. 
We have also investigated temperature variations of the thermodynamic functions and magnetocaloric effects inside different phases. In both systems the entropy increases by increasing temperature, while depending on the strength of magnetic field, the specific heat has a single-, double- and triple-peak structure. It increases by increasing temperature, shows a narrow peak in the supersolid and superfluid phases, and a broaden beak in the thermal solids and thermal insulator, and eventually decreases toward zero at higher temperatures.

%%%%%%%%%%%%%%%%%%%%%%%%%%%%%%%%%%%%%%%%%%%%%%%%%%
\ack
The authors would like to thank Andreas Honecker for introducing some references.

%%%%%%%%%%%%%%%%%%%%%%%%%%%%%%%%%%%%%%%%%%%%%%%%%%

\appendix
\section{}\label{appendix-A}
In this appendix we explain the details of the cluster mean field theory. In this method clusters of different sizes are considered on a sublattice background, so that interactions within clusters are included exactly and interactions with outsides are considered as effective fields \cite{heydarinasab2017inhomogeneous, heydarinasab2018spin}. Therefore one can partially take into account fluctuations around classical ground state as well as the effects of correlations of particles. In this respect the Hamiltonian of the system would be written as below:
\begin{equation}
H_{\rm{CMFT}}=H_C+\sum_{i\in
C}(\vec{h}_i^{eff}\cdot\vec{\s}_i+\vec{g}_i^{eff}\cdot\vec{\t}_i)
\label{CMFT-Hamiltonian},
\end{equation}
where the interactions within cluster $C$ are given by $H_C$, that contains the Hamiltonian in Eq. (\ref{eq:SMS-Hamiltonian}) and (\ref{eq:CAS-Hamiltonian}) with $i,j \in C$. While the interactions of spins inside the cluster with the rest of the system are included via the effective fields $\vec{h}_i^{eff}$ and $\vec{g}_i^{eff}$ for the subsystems with spins $\s$ and $\t$ respectively. The effective fields for the SMS model are:
\begin{eqnarray}
\nonumber\vec{h}_i^{eff}&=&\sum_{\la i,j\ra , j\in \bar{C}}[-2J(M_j^x\hat{x}+M_j^y\hat{y})+V_1M_j^z\hat{z}]\\
\nonumber &+& V_2\sum_{\la \la i,j\ra \ra , j\in \bar{C}} m_j^z\hat{z},\\
\nonumber\vec{g}_i^{eff}&=&\sum_{\la i,j\ra , j\in \bar{C}}[-2J(m_j^x\hat{x}+m_j^y\hat{y})+V_1m_j^z\hat{z}]\\
&+& V_2\sum_{\la \la i,j\ra \ra , j\in \bar{C}} M_j^z\hat{z}.
\end{eqnarray}
The effective fields for the CAS model are:
\begin{eqnarray}
\nonumber\vec{h}_i^{eff}&=&\sum_{\la i,j\ra , j\in \bar{C}}[-2J(m_j^x+M_j^x)\hat{x}-2J(m_j^y+M_j^y)\hat{y}\\
&+&\frac{V_1}{2} (m_j^z+M_j^z)\hat{z}]
\nonumber + V_2\sum_{\la \la i,j\ra \ra , j\in \bar{C}} M_j^z\hat{z},\\
\nonumber\vec{g}_i^{eff}&=&\sum_{\la i,j\ra , j\in \bar{C}}[-2J(m_j^x+M_j^x)\hat{x}-2J(m_j^y+M_j^y)\hat{y}\\
&+&\frac{V_1}{2}(m_j^z+M_j^z)\hat{z}]+ V_2\sum_{\la \la i,j\ra \ra ,
j\in \bar{C}} m_j^z\hat{z}.
\end{eqnarray}
In these equations $\bar{C}$ is the part of the system outside the
cluster. Also the magnetizations $\vec{m}_j=\la
\vec{\s}_j\ra$ and $\vec{M}_j=\la
\vec{\t}_j\ra$ are the expectation values within the CMFT which act as the mean fields on the spins $\s$ and $\t$. The order parameters $m_j^{x,y,z}$ and $M_j^{x,y,z}$ are calculated self-consistently as the expectation value of the spins inside the cluster.

At the zero temperature these expectation values are
calculated on the ground state of the system, and self-consistent
solutions should be done until the minimal ground state of the system
would be achieved. However at the finite temperatures $T$ all the ground
state and excited states of the system are contributed in the
solution and thermodynamic averages of the order parameters are calculated as:
%%%%%%%%%%%%%%%%%%
\begin{eqnarray}
\label{eq:CMFT-OP}
\no && m_j^{x,y,z}(T,h) = \langle \s^{x,y,z} \rangle = \frac{1}{\mathcal{Z}} Tr(\s e^{-H_{\rm{CMFT}}/K_{\rm{B}} T}), \\
\no && M_j^{x,y,z}(T,h) = \langle \t^{x,y,z} \rangle = \frac{1}{\mathcal{Z}} Tr(\t e^{-H_{\rm{CMFT}}/K_{\rm{B}} T}), \\
&& \mathcal{Z} = Tr(e^{-H_{\rm{CMFT}}/K_{\rm{B}} T}),
\end{eqnarray}
%%%%%%%%%%%%%%%%%%
where $K_{\rm{B}}$ is the Boltzmann constant. $H_{\rm{CMFT}}$ and $\mathcal{Z}$ respectively are corresponding CMFT Hamiltonian in Eq. (\ref{CMFT-Hamiltonian}) and partition function of the system. At the finite $T$, the free energy of the system i.e.
$F=\frac{1}{N}k_{\rm B}T \ln \mathcal{Z}$, should be minimized,
which $F$ is the free energy of the system. 
Finally the set of the CMFT energies that minimizes the free energy of the system, should be used for calculating any averages as the following equation:
%%%%%%%%%%%%%%%%%%
\begin{eqnarray}
\label{eq:CMFT-ave}
\langle A \rangle = \frac{1}{\mathcal{Z}} Tr(A e^{-H_{\rm{CMFT}}/K_{\rm{B}} T}),
\end{eqnarray}
%%%%%%%%%%%%%%%%%%
where $A$ is the corresponding function.

%%%%%%%%%%%%%%%%%%%%%%%%%%%%%%%%%%%%%%%%%%%%%%%%%%

\section{}\label{appendix-B}
In this appendix we explain how to obtain Eq. (\ref{eq:MCE3}) from Eq. (\ref{eq:MCE2}). We have:
%%%%%%%%%%%%%%%%%%
\begin{eqnarray}
\label{eq:B1}
\no \frac{\partial m}{\partial T}=\frac{\partial}{\partial T} \langle S^z \rangle&=&\frac{\partial}{\partial T} [ \frac{1}{\mathcal{Z}} Tr(S^z e^{-H/K_{\rm{B}} T})]\\
\no &=& \frac{Tr(S^z~ \partial e^{-H/K_{\rm{B}} T}/\partial T ) \mathcal{Z} - \partial \mathcal{Z}/\partial T ~Tr(S^z e^{-H/K_{\rm{B}} T})}{\mathcal{Z}^2}\\
\no &=& \frac{1}{k_{\rm B}T^2} [\frac{Tr(S^z H e^{-H/K_{\rm{B}} T})}{\mathcal{Z}} - \frac{Tr(H e^{-H/K_{\rm{B}} T}) Tr(S^z e^{-H/K_{\rm{B}} T})}{\mathcal{Z}^2}]\\
 &=& \frac{\langle H S^z \rangle -\langle H \rangle \langle S^z \rangle}{k_{\rm B}T^2}.
\end{eqnarray}
%%%%%%%%%%%%%%%%%%
Using above relations and inserting specific heat from Eq. (\ref{eq:S-Cv}), simply Eq. (\ref{eq:MCE3}) would be concluded.

%%%%%%%%%%%%%%%%%%%%%%%%%%%%%%%%%%%%%%%%%%%%
%%%%%%%%%%%%%%%%%%%%%%% References %%%%%%%%%%%%%%%%%%%%%%%%%%%%%%%%%%%%
\section*{References}
\bibliographystyle{iopart-num}
\bibliography{JPCM-15Nov}% Produces the bibliography via BibTeX.

\end{document}